\documentclass[twocolumn]{aastex62}
\usepackage{hyperref}
\hypersetup{bookmarksnumbered=true, bookmarksopen=true}

\usepackage{amssymb,amsmath,amsfonts}
\usepackage{newtxtext, newtxmath}
\usepackage{footnotebackref}

\allowdisplaybreaks[1]  

\usepackage{bm}
\usepackage{booktabs}
\usepackage{xcolor,colortbl}  
\usepackage{paralist}

\graphicspath{{figs/}}

\newcommand{\refsec}[1]{Section~\ref{#1}}
\newcommand{\reffig}[1]{Figure~\ref{#1}}
\newcommand{\reftab}[1]{Table~\ref{#1}}
\newcommand{\refeqn}[1]{Equation~(\ref{#1})}

\newcommand{\rperi}{r_{\mathrm{peri}}}

\renewcommand{\vr}{v_{\mathrm{r}}}
\newcommand{\vt}{v_{\mathrm{t}}}

\newcommand{\Robsmax}{R_{\mathrm{obs,max}}}

\newcommand{\rs}{r_{\mathrm{s}}}
\newcommand{\vs}{v_{\mathrm{s}}}

\newcommand{\rhos}{\rho_{\mathrm{s}}}

\newcommand{\msun}{M_{\odot}}

\newcommand{\kpc}{\mathrm{kpc}}
\newcommand{\kms}{\mathrm{km \, s}^{-1}}

\newcommand{\masyr}{\mathrm{mas \, yr}^{-1}}

\shorttitle{MILKY WAY MASS FROM PHASE-SPACE DISTRIBUTION OF SATELLITES}
\shortauthors{Li et al.}

\begin{document}

\title{\large\textbf{Constraining the Milky Way Mass Profile with Phase-Space Distribution of Satellite Galaxies}}

\author[0000-0001-7890-4964]{Zhao-Zhou Li}
\affil{Department of Astronomy, School of Physics and Astronomy,
  Shanghai Jiao Tong University, 800 Dongchuan Road, Shanghai 200240, China;
  \href{mailto:lizz.astro@gmail.com}{lizz.astro@gmail.com}, \href{mailto:ypjing@sjtu.edu.cn}{ypjing@sjtu.edu.cn}}

\author[0000-0002-3146-2668]{Yong-Zhong Qian}
\affil{School of Physics and Astronomy, University of Minnesota, Minneapolis, MN 55455, USA; 
  \href{mailto:qianx007@umn.edu}{qianx007@umn.edu}}
\affil{Tsung-Dao Lee Institute, Shanghai Jiao Tong University, 800 Dongchuan Road, Shanghai 200240, China}

\author[0000-0002-8010-6715]{Jiaxin Han}
\affil{Department of Astronomy, School of Physics and Astronomy,
  Shanghai Jiao Tong University, 800 Dongchuan Road, Shanghai 200240, China;
  \href{mailto:lizz.astro@gmail.com}{lizz.astro@gmail.com}, \href{mailto:ypjing@sjtu.edu.cn}{ypjing@sjtu.edu.cn}}
\affil{Kavli IPMU (WPI), UTIAS, The University of Tokyo, Kashiwa, Chiba 277-8583, Japan}

\author[0000-0002-9110-6163]{Ting S. Li}
\affil{Observatories of the Carnegie Institution for Science, 813 Santa Barbara St., Pasadena, CA 91101, USA}
\affil{Department of Astrophysical Sciences, Princeton University, Princeton, NJ 08544, USA}
\affil{NHFP Einstein Fellow}

\author{Wenting Wang}
\affil{Department of Astronomy, School of Physics and Astronomy,
  Shanghai Jiao Tong University, 800 Dongchuan Road, Shanghai 200240, China;
  \href{mailto:lizz.astro@gmail.com}{lizz.astro@gmail.com}, \href{mailto:ypjing@sjtu.edu.cn}{ypjing@sjtu.edu.cn}}
\affil{Kavli IPMU (WPI), UTIAS, The University of Tokyo, Kashiwa, Chiba 277-8583, Japan}

\author[0000-0002-4534-3125]{Y. P. Jing}
\affil{Department of Astronomy, School of Physics and Astronomy,
  Shanghai Jiao Tong University, 800 Dongchuan Road, Shanghai 200240, China;
  \href{mailto:lizz.astro@gmail.com}{lizz.astro@gmail.com}, \href{mailto:ypjing@sjtu.edu.cn}{ypjing@sjtu.edu.cn}}
\affil{Tsung-Dao Lee Institute, Shanghai Jiao Tong University, 800 Dongchuan Road, Shanghai 200240, China}

\begin{abstract}
We estimate the Milky Way (MW) halo properties using satellite kinematic data 
including the latest measurements from {\it Gaia} DR2. With a simulation-based 6D phase-space 
distribution function (DF) of satellite kinematics, 
we can infer halo properties efficiently and without bias, and handle the selection function
and measurement errors rigorously in the Bayesian framework. Applying our DF from the EAGLE simulation 
to 28 satellites, we obtain an MW halo mass of $M=1.23_{-0.18}^{+0.21}\times 10^{12} M_\odot$ and 
a concentration of $c=9.4_{ -2.1}^{ +2.8}$ with the prior based on the $M$-$c$ relation. 
The inferred mass profile is consistent with previous 
measurements but with better precision and reliability due to the improved methodology and data. 
Potential improvement is illustrated by combining satellite data and stellar rotation curves. 
Using our EAGLE DF and best-fit MW potential, we provide much more precise estimates 
of kinematics for those satellites with uncertain measurements.
Compared to the EAGLE DF, which matches the observed satellite kinematics very well, 
the DF from the semi-analytical model based on the dark-matter-only simulation Millennium II
(SAM-MII) over-represents satellites with small radii and velocities. We attribute this difference 
to less disruption of satellites with small pericenter distances in the SAM-MII simulation.
By varying the disruption rate of such satellites in this simulation, we estimate
a $\sim 5\%$ scatter in the inferred MW halo mass among hydrodynamics-based simulations.
\end{abstract}


\section{Introduction} \label{sec:intro}
The total mass and density distribution for the Milky Way (MW) dark matter halo are
of great importance to various astrophysical studies.
Most of the methods that have been proposed to constrain the MW mass profile make use of dynamical tracers
(see \citealt{Wang2019} for review).
The mass distribution of the inner halo (within $\sim 40\,\kpc$) 
has been relatively well constrained by
the kinematics of masers, stars, stellar streams, and globular clusters.
However, the profile of the outer halo and the virial mass show less agreement
(see \citealt{Eadie2018,Wang2019} for comparisons of recent estimates,
and \citealt{McMillan2011} for critical comments).
Due to the limited number of tracers and lack of good data, different model assumptions
(including profile extrapolation) and their associated systematics lead to
a factor of $\gtrsim 2$ disagreement in the halo mass estimate.

Satellite galaxies are the preferred tracers for the outer halo in several aspects.
First of all, thanks to their relatively high luminosities and extended spatial distribution,
currently they are the only tracers with sufficient statistics for the very outer halo ($\gtrsim 100$ kpc).
Further, their kinematics is well understood in the framework of
hierarchical structure formation and accurately modeled by modern cosmological simulations,
which makes their dynamical modeling more reliable.
In addition, satellite galaxies closely trace the underlying phase-space distribution of dark matter particles, while halo stars are less phase-mixed \citep{Han2019}.

A popular method for dynamical modeling of outer halo tracers
is based on the phase-space distribution function (DF).
As the complete statistical description of a stationary dynamical system,
the DF can maximize the use of kinematic data.
The DF method has been widely used for tracers like stars, globular clusters, and satellite galaxies
(e.g., \citealt{Little1987,Kochanek1996,Wilkinson1999,Sakamoto2003,Deason2012,Williams2015a,Binney2017,Eadie2018,Posti2018,Vasiliev2018}).
However, despite many analytical and simulation-based attempts
(e.g., \citealt{Cuddeford1991,Evans2006,Wojtak2008,Posti2015,Williams2015}) since the seminal work of \citet{Lynden-Bell1967}, 
an accurate and explicit form of the DF for tracers of halos remains to be found and verified.
As shown by \citet{Wang2015b} and \citet{Han2016a},
unjustified assumptions (e.g., constant velocity anisotropy)
in constructing the DF may lead to substantially biased results.
Fortunately, we can construct the DF for satellite galaxies directly from cosmological simulations.

\citet{Li2017} constructed the probability density function (PDF) $p(E,L)$ of the satellite orbital energy $E$ 
and angular momentum $L$ directly from cosmological simulations. They found that the internal dynamics 
of different halos are very similar when normalized by the corresponding virial scales. Using this feature and the
constructed $p(E,L)$, they developed a method to estimate the halo mass from satellite kinematics.
\citet{Callingham2018} made some improvement of this method.
With the kinematic data of 10 luminous satellites, they found an MW halo mass of 
$1.17^{+0.21}_{-0.15} \times 10^{12} \msun$.
However, the PDF $p(E,L)$ in the 2D orbital space of $E$ and $L$ differs from the DF $f({\bm{r}},{\bm{v}})$ 
in the 6D phase space of position $\bm{r}$ and velocity $\bm{v}$. Because the orbital energy $E$ is not directly 
observable, the use of $p(E,L)$ to estimate the halo mass requires calibration with mock samples. In contrast,
as shown in \citet{Li2019}, the use of the DF $f({\bm{r}},{\bm{v}})$, which describes the direct 
observables ${\bm{r}}$ and ${\bm{v}}$, automatically gives unbiased and precise estimates of halo properties.
The precision of this DF method can be attributed to the incorporation of both the orbital distribution described
by $p(E,L)$ and the radial distribution along each orbit that was the basis of the orbital PDF method \citep{Han2016b}.

Assuming steady state for satellites in the host halo potential,
\citet{Li2019} used both the similarity of the internal dynamics for different halos and
the universal Navarro--Frenk--White (NFW, \citealt{Navarro1996}) density profile in constructing
the DF from a cosmological simulation. Consequently, they were able to estimate both the halo mass $M$
and the concentration $c$ for the NFW profile, thereby obtaining the mass distribution. 
Tests with mock samples showed that this method is valid and accurate, as well as
more precise than pure steady-state methods, including the Jeans equation and Schwarzschild modeling.
The halo-to-halo scatter due to diversities in halo formation history and environment
results in an intrinsic uncertainty of only $\sim 10\%$ for the halo mass.
In addition, this method facilitates a rigorous 
and straightforward treatment of various observational effects, including selection 
functions and observational errors. This feature is especially important for outer halo tracers,
for which these effects are much more severe and their improper treatment can lead to serious bias.

In this paper, we apply the DF method of \citet{Li2019} to estimate the MW halo properties
using kinematic data on 28 satellites, 
including precise proper motion measurements by {\it Gaia} DR2 \citep{GaiaCollaboration2018a}.
This sample is optimized for the outer halo.
The improved methodology and observational data enable us to obtain the currently best estimates of the 
MW halo mass and outer halo profile. Our results weakly depend on the simulation used to construct the DF.
We quantify this model dependence by comparing the results from 
the hydrodynamics-based EAGLE simulation  and the semi-analytical model based on the dark-matter-only simulation
Millennium II. We confirm by the goodness-of-fit that
the EAGLE simulation provides a better description of the kinematics of MW satellites.

The plan of this paper is as follows. 
We describe the satellite sample and the corresponding selection function in \refsec{sec:data}.
We outline our method in \refsec{sec:method} and
present the results in \refsec{sec:result}.
We make comparisons with previous works 
and show how our results can be improved by combining different tracer populations in \refsec{sec:comp_joints}.
We summarize our results and give conclusions in \refsec{sec:conclusion}.

In this paper, the halo mass $M$ and concentration $c$ refer to the \textit{total} mass
including the baryonic contribution. We define $M$ as the mass enclosed by the virial radius $R$,
within which the average density is 200 times the critical density of the present universe,
$\rho_{\rm cri}={3H_0^2}/({8\pi G})$. Here,
$H_0=67.77\,\mathrm{km\,s^{-1}Mpc^{-1}}$ is the Hubble constant and $G$ is the gravitational constant.

\section{Observation data} \label{sec:data}

In this work, we use the recent MW satellite data,
including the coordinates, luminosities, distances, line-of-sight velocities, and proper motions,
compiled by \citet{Riley2018}.
When available, we adopt the ``gold" proper motions in \citet{Riley2018}, which usually represent more precise measurements due to the larger sample of member stars used.
Furthermore, this compilation omitted
satellites that have been disrupted or whose nature is still under debate.

\subsection{Satellite sample} \label{sec:satellites}

We select our sample of satellites based on their distance to the Galactic center (GC), $r$.
Considering $r=262\pm 9\, \kpc$ for Leo I, the farthest satellite with measured proper motion in \citet{Riley2018},
we only use those satellites with $r<280$ kpc.
Varying this upper limit within 100--300 $\kpc$ does not change our results (see \refsec{sec:robust}).
We also exclude satellites with $r< 40\,\kpc$ to avoid complications from the MW disk.
Based on the above criteria ($40<r<280\,\kpc$), we have selected 28 satellites, whose properties are 
listed in \reftab{tab:properties} of Appendix \ref{sec:data_table}. 
The median distance to the sun for this sample is $\sim 100$ kpc.

As our model uses kinematic data relative to the GC but the satellite data are given in
the Heliocentric Standard of Rest (HSR) frame, we transform the HSR data (coordinates, distance, 
line-of-sight velocity, and proper motion) to quantities in the Galactocentric Standard of Rest (GSR)
frame with the Python package \texttt{Astropy} (\citeyear{AstropyCollaboration2013}).
We adopt the following position and velocity of the Sun
in the GSR frame \citep{Bland-Hawthorn2016a}: a radial distance of $8.2\,\kpc$ in the Galactic plane,
a vertical distance of $25\,\mathrm{pc}$ above this plane, and
$(U_{\odot}, V_{\odot}, W_{\odot}) = (10, 248, 7)\,\kms$,
where $U_{\odot}$ is the velocity toward the GC, 
$V_{\odot}$ is positive in the direction of Galactic rotation, and
$W_{\odot}$ is positive toward the north Galactic pole.
Measurement errors in the distance, line-of-sight velocity, and proper motion are taken into account as follows.
Assuming that the error in each observable is Gaussian and mutually independent,
we generate 2000 Monte Carlo realizations of the HSR data for each satellite according to these errors
and transform each realization to propagate the errors to the GSR data.
The GSR data thus obtained will be used as the direct input for our model.

\subsection{Selection function} \label{sec:selec_func}

Satellite samples discovered by sky surveys inevitably suffer from incompleteness
[see e.g., \citealt{Koposov2008,Walsh2009} for Sloan Digital Sky Survey (SDSS) and \citealt{Jethwa2016} for Dark Energy Survey (DES)].
Here, we only use a subset of MW satellites with complete astrometric data measured by {\it Gaia} DR2,
which represents a uniform survey with a well-understood selection function.
Below we derive a good approximation of this selection function.

\begin{figure}[btp]
  \centering
  \includegraphics[width=0.45\textwidth]{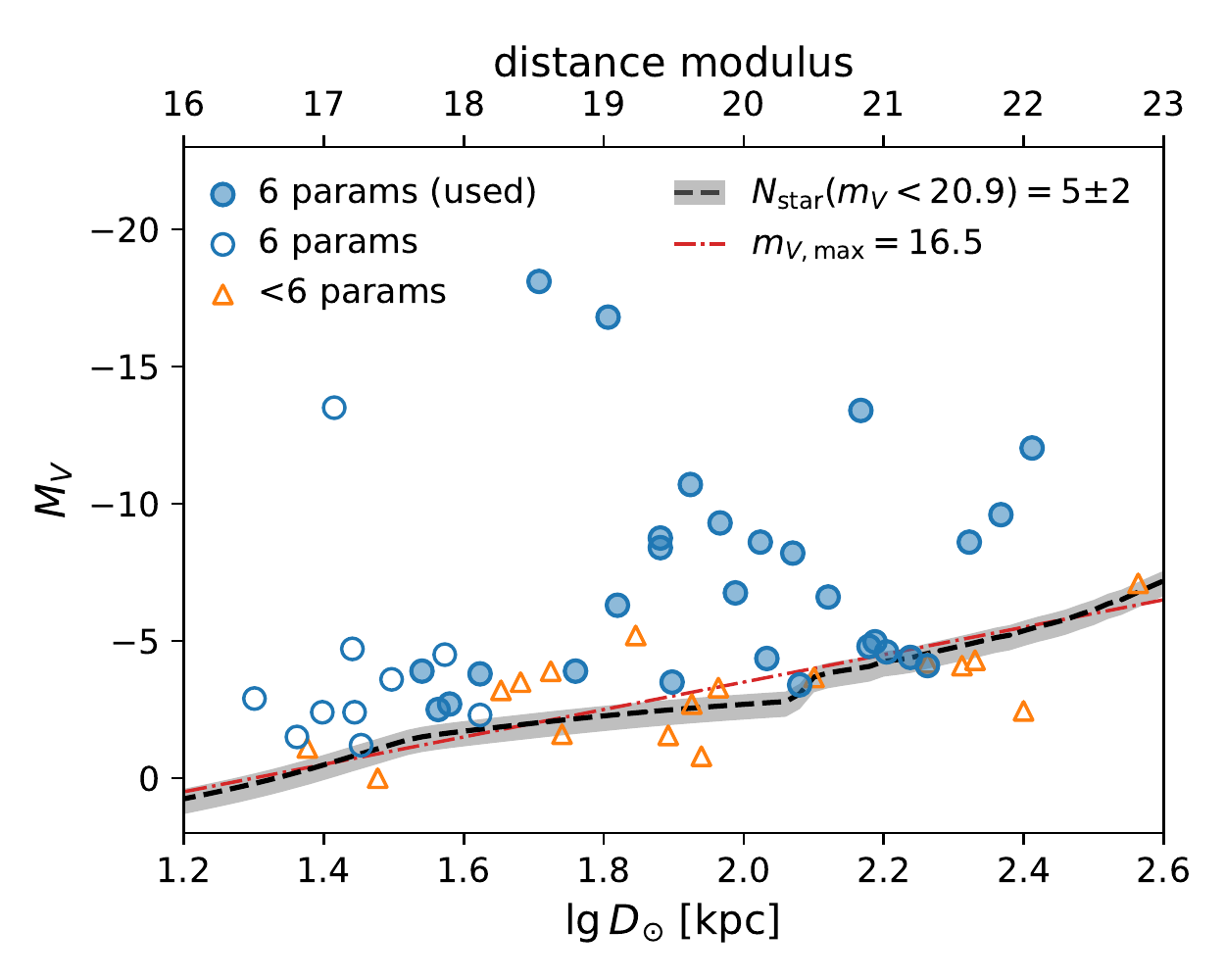}
  \caption{
    Heliocentric distance $D_\odot$ and absolute magnitude $M_V$ for known MW satellites.
    Satellites with complete kinematic data \citep{Riley2018} measured by {\it Gaia} are shown as circles
    (labeled ``6 params'' for 6 parameters),
    and the rest (taken from \citealt{Simon2019}) are shown as triangles.
    The 28 satellites with $40<r<280\,\kpc$ used in this work are marked by filled blue circles.
    The black dashed curve shows the completeness distance $\Robsmax$ for {\it Gaia} proper motion
    measurement at each $M_V$, which distinguishes the circles and triangles very well.
    The red dash-dotted line indicates an apparent magnitude of $m_{V\!,\max}=16.5$ for reference.
    See text for details.
  }
  \label{fig:obs_selection}
\end{figure}

{\it Gaia} DR2 can usually measure the proper motion of a satellite reliably only when 
it contains at least $\sim 5$ member stars brighter than the observation limit
(apparent $G$-band magnitude of $m_G \!\sim\! 20.9$ for\defcitealias{GaiaCollaboration2018a}{DR2}\citetalias{GaiaCollaboration2018a}).
The detection rate drops sharply below this threshold (see black dashed curve in \reffig{fig:obs_selection}).
Therefore, we can estimate the completeness radius $\Robsmax$ as a function of the satellite luminosity.
For each satellite, we generate a number of synthetic galaxies according to its $V$ band luminosity $M_V$.
For each synthetic galaxy, a single stellar population is simulated
with a \citet{Chabrier2001} mass function, a typical age of 12.5 Gyr, and a metallicity of $\mathrm{[Fe/H]} = -2.2$ 
using the PARSEC isochrone online library\footnote{ \ \url{http://stev.oapd.inaf.it/cgi-bin/cmd_3.2}}
\citep{Bressan2012}.
We then determine the $\Robsmax$ at which the synthetic galaxies
contain an average of $N_{\rm star}=5$ stars brighter than $m_V=20.9$.
As shown by the black dashed curve\footnote{
At $D_\odot\sim 120$ kpc corresponding to a distance modulus of $\sim 20.4$,
horizontal branch stars ($M_V \sim 0.5$) become too dim and the number of dwarf satellites 
accessible to {\it Gaia} astrometry drops markedly. This effect gives rise to the abrupt change
at $D_\odot\sim 120$ kpc in $\Robsmax$.
}
in \reffig{fig:obs_selection}, the $\Robsmax$ derived above 
distinguishes the satellites with and without complete kinematic data very well.\footnote{
Four satellites (Pictor II, Tucana IV, Grus II, Sagittarius II) are clearly within their completeness radii but do not have complete kinematic data. 
These satellites are accessible through current surveys and facilities, but 
kinematic measurements were either incomplete or unavailable to us when we started this study.
Note that the proper motions for three of them were recently published
(Sagittarius II by \citealt{2020MNRAS.491..356L};
Grus II and Tucana IV by \citealt{2019arXiv191108493S}).
Because the absence of the above four satellites from our sample is unrelated to their kinematics,
ignoring them does not affect our analysis.
}
The shaded band around this curve corresponds to $N_{\rm star}=5\pm 2$.
It is close to a cut of $m_{V\!,\max}=16.5$ (red dash-dotted line) for the total apparent magnitude of satellites,
i.e., $\Robsmax(M_V) = 10^{- 0.2 (M_V-m_{V\!,\max}) - 2}\,\kpc$.
It is important to use the appropriate DF within the $\Robsmax$ for each satellite [see Equation (\ref{eqn:selection})].
Otherwise, the halo concentration can be seriously overestimated (see \refsec{sec:robust}).

In addition to the selection on distance,
the spatial distribution of satellites is further affected by the angular coverage of
the sky surveys that discovered them.
This effect is especially severe for the low Galactic latitude ($|b|<15^{\circ}$) region,
which is blocked by dense dust and disk stars in the foreground (see discussion in \citealt{Torrealba2018}).
However, this angular selection does not affect our analysis under the assumption of spherical symmetry. 

A special class of satellites, the ultra diffuse dwarfs (e.g., surface brightness fainter than $30\, \mathrm{mag/arcsec^2}$ for SDSS),
merit discussion. They can actually have very high total luminosities, but are inaccessible to current satellite searching algorithms \citep{Koposov2008}.
For example, the recently discovered Antlia 2 (first identified using astrometry data from {\it Gaia})
has $M_V = -8.5$ but a very low surface brightness of 
$32.3\, \mathrm{mag/arcsec^2}$ \citep{Torrealba2018}.
\citet{Li2019} showed that kinematics of satellites is largely independent of their luminosities.
It seems reasonable to assume that kinematics is also independent of surface brightness.
In this case, absence of ultra diffuse dwarfs in our sample does not affect our analysis, either.
Nevertheless, the effects of such satellites on the DF method warrants further studies.

\section{Method}\label{sec:method}

We briefly describe our method in this section. 
Much more detail can be found in \citet{Li2019},
where the method was carefully tested for its validity and performance 
with MW-like halos from a cosmological simulation.

\subsection{Simulation-based DF}\label{sec:method-DF}

We construct the DF for satellites of MW-like halos from cosmological simulations
based on the following assumptions:

(1) All halos have the spherical NFW density profile \citep{Navarro1996},
\begin{equation}
\rho(r)=\frac{\rhos}{(r/\rs)(1+r/\rs)^2},
\end{equation}
where $\rhos$ and $\rs$ are the characteristic density and radius, respectively.
A specific set of $\rhos$ and $\rs$ corresponds to a specific set of halo mass $M=\int_0^R4\pi\rho(r)r^2dr$ 
and concentration $c=R/\rs$, where $R$ is the virial radius.

The NFW profile is known to give a good description of halos in dark-matter-only simulations.
Here, we apply it to the total density including the baryonic contribution.
Whereas simulations have not reached consensus on the influence of baryonic processes in the inner halo,
they agree that the outer halo dominated by dark matter is less affected by these processes (e.g., \citealt{Schaller2015,Kelley2018})
and is well described by the NFW profile for $r\gtrsim 0.05R$ \citep{Schaller2015}.

(2) The satellites are in dynamical equilibrium
with their host halo, so their kinematics 
in terms of $\bm{r}$ and $\bm{v}$ can be described by a 
steady-state DF in phase space
\begin{equation}
\frac{d^6N}{d^3{\bm{r}}d^3{\bm{v}}}=f({\bm{r}},{\bm{v}}).
\label{eqn:frv}
\end{equation}
Note that the velocity distribution of satellites is largely unchanged by baryonic physics
for $r \gtrsim 0.25R$ \citep{Sawala2017,Richings2018}.

(3) The internal dynamics of all halos are similar after $\bm{r}$ and 
$\bm{v}$ are normalized by their characteristic scales, 
$\rs$ and $\vs=\rs\sqrt{4\pi G\rhos}$\,,
respectively. Therefore, the dimensionless DF $\tilde f(\tilde{\bm{r}},\tilde{\bm{v}})$ 
in terms of the dimensionless variables 
$\tilde{\bm{r}}={\bm{r}}/\rs$ and $\tilde{\bm{v}}={\bm{v}}/\vs$ is universal to all halos.
For a halo of mass $M$ and concentration $c$, the DF of its satellites is
\begin{equation}
f(\bm{w}|M,c)\equiv f (\bm{r}, \bm{v}|M, c) = 
    \frac{1}{\rs^3 \vs^3} \tilde{f} \left(\frac{{\bm{r}}}{\rs}, \frac{{\bm{v}}}{\vs} \right),
\end{equation}
where $\bm{w}$ denotes the set of $\bm{r}$ and $\bm{v}$ for a satellite.

The general validation of the above assumptions is presented in \citet{Li2019}.
Nevertheless, individual halos are still expected to exhibit certain deviations from these assumptions
due to diversities in their formation histories and environments.
As shown in \citet{Li2019} and \refsec{sec:main_result},
the consequent systematic uncertainty can be quantified 
by tests with realistic mock samples.

We construct the universal dimensionless DF $\tilde{f}(\tilde{\bm{r}},\tilde{\bm{v}})$
by stacking template halos in a cosmological simulation (see \citealt{Li2019} for details).
Under our assumptions, $\tilde{f}(\tilde{\bm{r}},\tilde{\bm{v}})=\tilde{f}(\tilde{E},\tilde{L})$,
which means that $\tilde{f}$ depends on $\tilde{\bm{r}}$ and $\tilde{\bm{v}}$
only through the form of the dimensionless energy $\tilde{E}=E/\vs^2$ and the dimensionless angular momentum
$\tilde{L}=L/(\rs\vs)$.
Compared to the conventionally adopted analytical DFs,
our simulation-based DF is more realistic and automatically treats unbound orbits.
Thus, we do not have to assume whether any satellite, e.g., Leo I, is bound to the MW or not.

\begin{figure*}[htbp]
  \centering
  \includegraphics[width=0.45\textwidth]{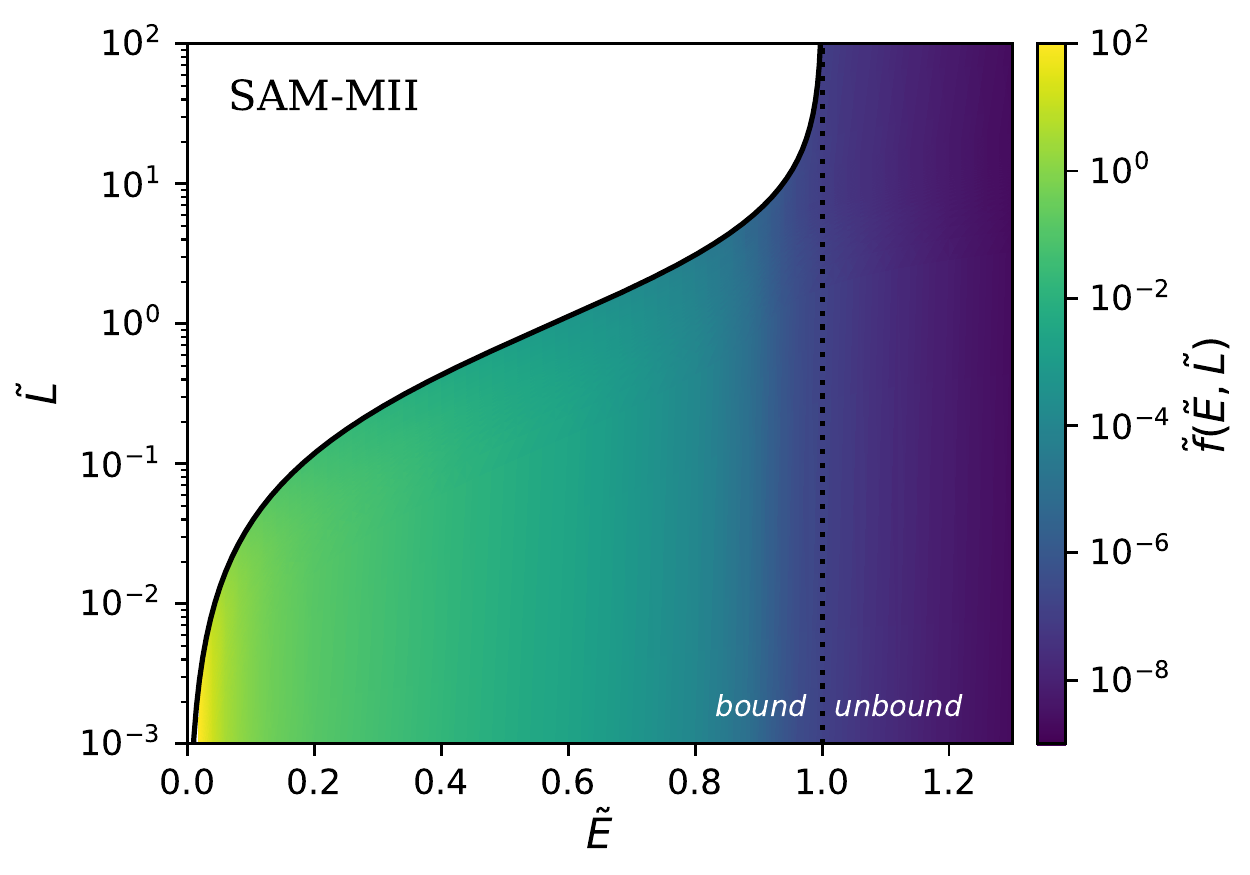}
  \includegraphics[width=0.45\textwidth]{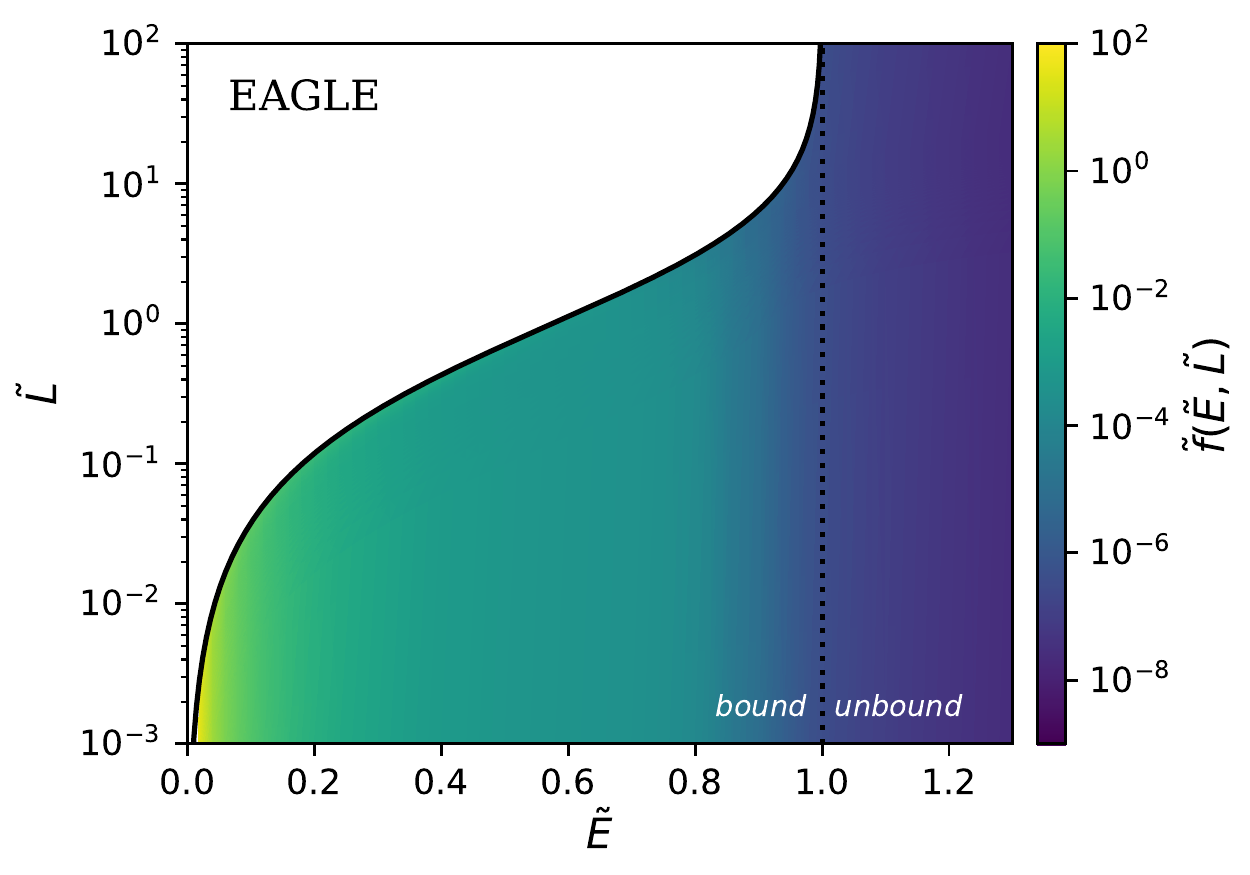}
  \caption{
    Dimensionless DFs
    $\tilde{f} (\tilde{E}, \tilde{L})$
    constructed from the SAM-MII (left) and EAGLE (right) simulations.
    The two DFs are quite similar,
    but the EAGLE DF has fewer tightly-bound (small $\tilde{E}$) satellites due to enhanced satellite disruption.
    The energy $\tilde{E}$ is always positive because the halo center is adopted as the zero point of the potential.
    Note that satellites with $\tilde{E}>1$ (to the right of the dotted line) are not bound.
    Note also that $f(E, L)$ is the DF in the 6D phase space of $\bm{r}$ and $\bm{v}$.
  }
  \label{fig:DF}
\end{figure*}

We expect that our DF has some dependence on the simulation used.
In this work, we use template halo samples from two distinct simulations.
For both samples, halos have the same mass range of $10^{11.5} \leq M/\msun \leq 10^{12.5}$,
and their luminous satellites within $25\rs$ are selected to construct the DF.
One sample, the same as used in \citet{Li2019}, is from the galaxy catalog generated by a semi-analytical model 
(SAM; \citealt{Guo2011}) based on the dark-matter-only simulation
Millennium II (MII; \citealt{Boylan-Kolchin2009}). This sample
contains $\sim\! 10^4$ isolated halos with a total of $\sim\! 10^5$ satellites, each having a stellar mass
of $m_\star \geq 100\,\msun$.
The other sample is selected from the hydrodynamics-based EAGLE Simulation \citep{Schaye2015,Crain2015,McAlpine2016}
with the same criteria as in \citet{Callingham2018} except for the halo mass range.
This sample contains $\sim\! 1700 $ relaxed halos with a total of $\sim\! 2.5\times 10^4$ satellites,
each having at least one star particle.
The concentration for the NFW profile of each halo is taken from \citet{Wang2017b} for the MII simulation and 
\citet{Schaller2015} for the EAGLE simulation.
For EAGLE halos, the concentration is fitted for the profile over $r=(0.05$--$1)R$.

The DFs $\tilde{f}(\tilde{E},\tilde{L})$ constructed from the SAM-MII and EAGLE simulations are
shown in Figure~\ref{fig:DF}. It can be seen that they are quite similar, but the EAGLE DF has
fewer tightly-bound (small $\tilde{E}$) satellites. Using the same tests 
as for the SAM-MII DF in \citet{Li2019}, we have checked that the EAGLE DF provides unbiased
estimates of the mass and concentration for the underlying halo sample. In particular, as discussed
in Appendix~\ref{sec:massive_test}, these estimates are insensitive to the presence of
a massive neighbor or satellite like M31 or the Large Magellanic Cloud (LMC), respectively, for the MW.
Comparing the results from the SAM-MII and EAGLE DFs allows us to assess the systematic uncertainties
due to the simulation used. We will also show by the goodness-of-fit in \refsec{sec:model_selec}
that the EAGLE simulation matches the observations better.

\subsection{Estimating halo properties} \label{sec:estimate}
Within the Bayesian statistical framework, we can use our DF to infer the halo mass $M$ and concentration $c$ 
efficiently and without bias. In addition, we can treat various observational effects,
including the selection function (incompleteness) and measurement errors,
in a rigorous and straightforward manner.

As discussed in \refsec{sec:data}, 
we consider only those MW satellites with $40 < r < 280\,\kpc$
and use an approximate selection function based on a luminosity-dependent
completeness radius $\Robsmax$ for each satellite.
Consequently, the PDF including the selection function is
\begin{equation}
  p_{\mathrm{s}} (\bm{w}|M, c) = 
    \frac{f (\bm{w} | M, c)} {\int_{r_{\min} < r' < \min\{r_{\max}, \Robsmax\}} f (\bm{w}' |M, c) d^6 \bm{w}'},
\label{eqn:selection}
\end{equation}
\phantom{}\par\noindent
where $r_{\min} = 40\,\kpc$ and $r_{\max} = 280\,\kpc$.\footnote{
Strictly speaking, $\Robsmax$ is a Heliocentric distance. However, because all the satellites in our
  sample are sufficiently far away from the GC, $\Robsmax$ can be taken as a Galactocentric distance 
  to good approximation.}
Note that under our assumption of spherical symmetry, an angular selection function
adds the same constant factor to both the numerator and the denominator, thus, having no effect
on the above DF.

We further take observational errors into account through the hierarchical Bayesian technique,
by integrating over all possible $\bm{w}$ corresponding to the observed $\hat{\bm{w}_i}$
for a satellite, to obtain
\begin{equation}
  p_{\mathrm{ob}} (\hat{\bm{w}}_i |M, c) 
    = \int p_{\mathrm{err}}(\hat{\bm{w}}_i | \bm{w})  p_{\mathrm{s}}(\bm{w} | M, c)  d^6 \bm{w},
\label{eqn:error}
\end{equation}
where $p_{\mathrm{err}} (\hat{\bm{w}} |{\bm{w}})$ describes 
the deviation of observables from their true values due to measurement errors.
In practice, we use the Monte Carlo integration method to simplify the above calculation
(e.g., \citealt{Callingham2018}).
Specifically, for each satellite labeled $i$, 
we generate Monte Carlo realizations $\{\bm{w}_{ik} \}_{k=1,\,2,\cdots}$ that follow
$p (\bm{w}_{ik}) \propto p_{\mathrm{err}} (\hat{\bm{w}}_i | \bm{w}_{ik})$ 
(see \refsec{sec:satellites})
and take the average of $p_{\mathrm{s}} (\bm{w}_{ik} |M, c)$ for these realizations.\footnote{
  The result of the Monte Carlo integration differs from the actual integration
  by a constant factor
  $\int p_{\mathrm{err}} (\hat{\bm{w}}_i | \bm{w}) d^6\bm{w}$,
  which is unity only for
  $p_\mathrm{err} (\bm{w}_1 |\bm{w}_2) = p_{\mathrm{err}} (\bm{w}_2 |\bm{w}_1)$. 
  However, such a constant factor does not affect any of the following analysis.}

Using an observed sample of $n_\mathrm{sat}$ satellites
with kinematic data $\{ \hat{\bm{w}}_i \}_{i=1,\ldots,n_\mathrm{sat}}$,
we can now infer the mass and concentration of the MW halo from the Bayesian formula
\begin{gather}
  p (M, c| \{ \hat{\bm{w}}_i \}) = \frac{1}{Z} 
  \left[ {\textstyle\prod\nolimits^{n_\mathrm{sat}}_{i = 1}}\, p_{\mathrm{ob}} ({\hat{\bm{w}}}_i |M, c) \right]
     p (c|M) p (M), \label{eqn:posterior}
\end{gather}
where $p (M)$ and $p (c|M)$ represent our prior knowledge, and
the normalization factor $Z$, also known as the Bayesian evidence,
is given by integrating (marginalizing) the posterior (i.e., the expression after $1/Z$) over 
all possible model parameters $M$ and $c$.
A model with a higher $Z$ is more favored as it gives a higher probability of obtaining the observational data
(see e.g., \citealt{Trotta2008}).

We use flat priors on both $\lg M$ and $\lg c$ by default to avoid relying on extra information.
Based on cosmological simulations, the concentration for halos of the same mass follows a log-normal distribution
with a scatter of $\sigma_{\lg c} \simeq 0.15\, \mathrm{dex}$ \citep{Jing2000a}.
Combining this result and the median $M$-$c$ relation derived from the EAGLE simulation \citep{Schaller2015}
gives an alternative prior on $\lg c$, which we take to be
\begin{equation}
    p(\lg c|M)=\mathcal{N} \left(0.912 - 0.087 \lg (M/10^{12} \msun), 0.15\right).
\label{eqn:cMR}
\end{equation}
Using the above prior or a similar one based on dark-matter-only simulations (e.g., \citealt{Dutton2014}) 
gives almost the same inferred halo properties.

\section{Results} \label{sec:result}

In this section, we apply our method to infer the MW halo
properties based on the SAM-MII and EAGLE DFs, respectively, 
using the kinematic data for our sample of 28 satellites.
We always use a flat prior on $\lg M$, but we present results
for both a flat prior on $\lg c$ and the alternative prior
[\refeqn{eqn:cMR}] based on the $M$-$c$ relation.

\subsection{Halo mass and concentration} \label{sec:main_result}

\begin{figure*}[htbp]
  \centering
  \includegraphics[width=0.45\textwidth]{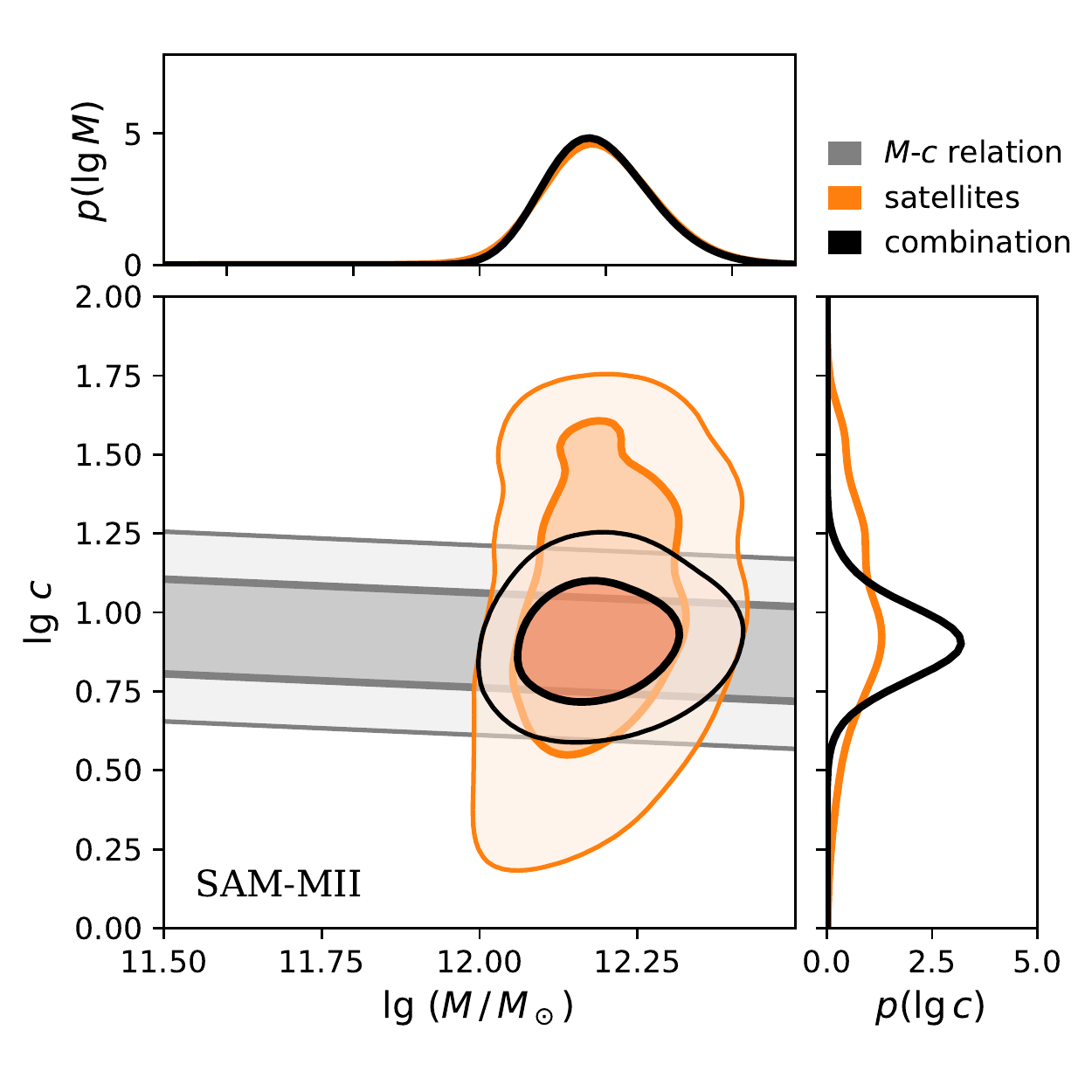}
  \includegraphics[width=0.45\textwidth]{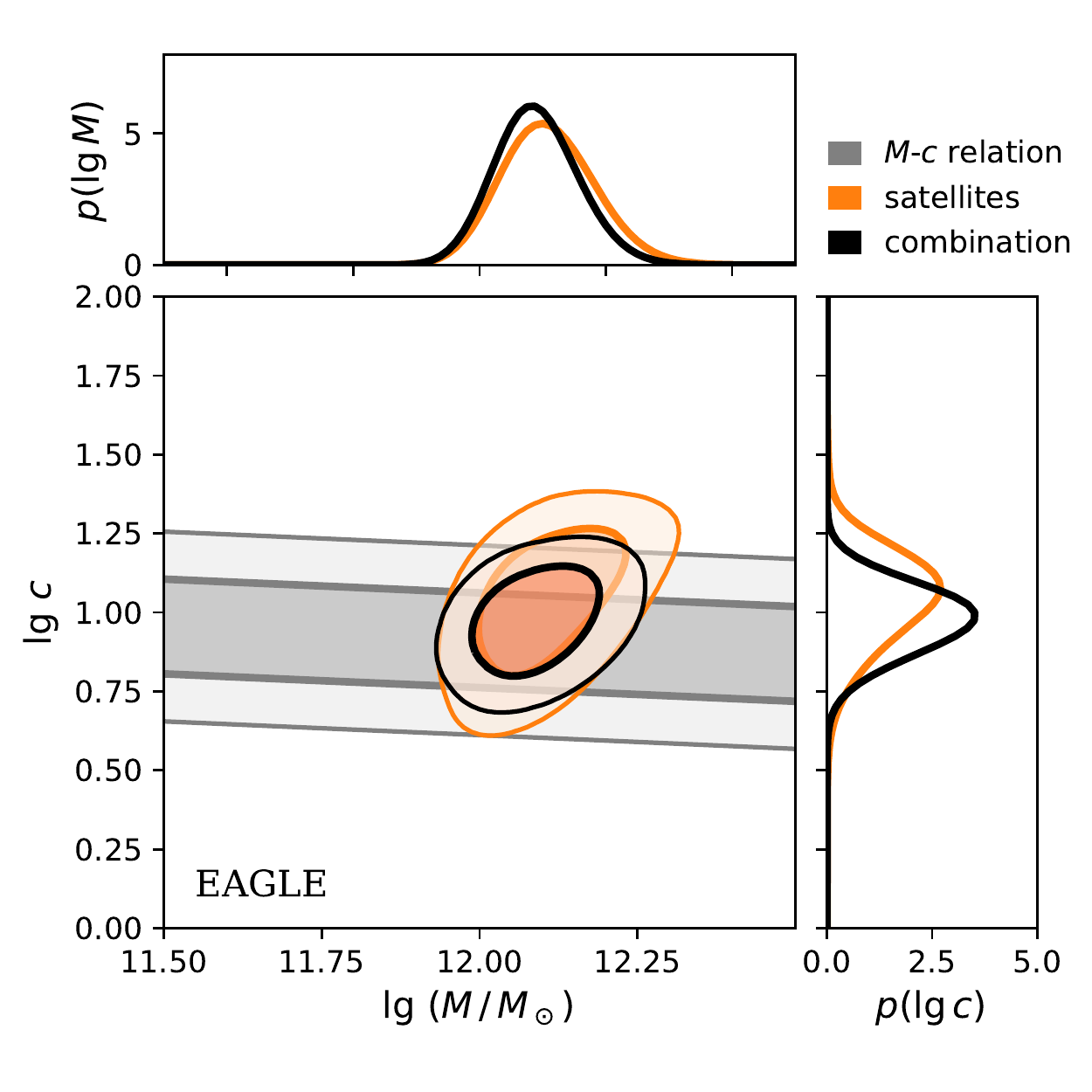}
  \vspace{-4mm}
  \caption{
  The MW halo mass and concentration inferred from the SAM-MII (left panel) and EAGLE (right panel) DFs.
  A flat prior on $\lg M$ is always used. Results are presented for both a flat prior on $\lg c$ (orange color) and
  the alternative prior (black color) based on the $M$-$c$ relation shown by the gray $1\sigma$ and $2\sigma$ contours.
  The other contours show the $1\sigma$ and $2\sigma$ confidence regions for the inferred $\lg M$ and $\lg c$.
  The corresponding marginalized distributions are shown in the upper and right subpanels.
  }
  \label{fig:esti_contour}
\end{figure*}

Using the flat priors on $\lg M$ and $\lg c$, we calculate their joint probability distribution 
from \refeqn{eqn:posterior} on a 2D grid. 
The inferred $1\sigma$ (68.3\%) and $2\sigma$ (95.4\%) confidence regions 
are shown in \reffig{fig:esti_contour}.
The marginalized distributions of $\lg M$ and $\lg c$ are shown in the top and right subpanels, 
respectively, from which the marginalized $1\sigma$ confidence interval is obtained for each parameter.
The above results are consistent with the $M$-$c$ relation. As expected, using the alternative prior
based on this relation significantly reduces the uncertainty in the estimated $\lg c$, especially 
for the SAM-MII DF. However, because the inferred $\lg M$ depends on $\lg c$ only weakly, using
the alternative prior improves the precision of the estimated $\lg M$ only slightly.

The best-fit values of $M$ and $c$ corresponding to the maximum posterior
are listed in \reftab{tab:esti_result} along with the marginalized $1\sigma$ uncertainties. 
Parameters for the 2D Gaussian fit to the joint probability distribution of $\lg M$ and $\lg c$,
including the mean and standard deviation of each variable and the correlation coefficient $\rho_\mathrm{corr}$,
are also provided there for convenience of use.
While the results from the SAM-MII and EAGLE DFs are consistent with each other at the $1\sigma$ level,
the inferred halo mass is larger and has a larger statistical uncertainty for the SAM-MII DF.
As shown in \refsec{sec:model_selec}, the hydrodynamics-based EAGLE DF is significantly favored 
over the SAM-MII DF by the observations.
Therefore, the results from the EAGLE DF are recommended.
The enclosed mass $M(<r)$ within radius $r$ inferred from this DF is
given for $r=30$--$400\,\kpc$ in Appendix~\ref{sec:profile_table}.

\begin{table}[tbp]
\caption{
  Inferred MW halo mass and concentration.
  Results are presented for the SAM-MII and EAGLE DFs for a flat prior on $\lg c$
  and the alternative prior based on the $M$-$c$ relation, as well as for combined
  tracer populations. Each set of results comprises the best-fit values of $M$ and $c$ 
  with $1\sigma$ uncertainties, and the parameters for
  the 2D Gaussian fit to the joint probability distribution of $\lg M$ and $\lg c$,
  including the mean and standard deviation of each variable 
  and the correlation coefficient $\rho_\mathrm{corr}$.
  The results from the EAGLE DF (in bold) are recommended.
}
\centering
\begin{tabular}{lrrrr}
\midrule \midrule 
    & \multicolumn{2}{c}{Satellites} & \multicolumn{2}{c}{Satellites + Halo Stars}\\
\cmidrule(lr){2-3} \cmidrule(lr){4-5}
    & {flat prior} & \multicolumn{1}{c}{$M$-$c$ relat.} & {flat prior} & \multicolumn{1}{c}{$M$-$c$ relat. }\\
\midrule
\multicolumn{3}{l}{SAM-MII}\\
\cmidrule(lr){1-5}
  $M^\text{a}$ & $1.55_{-0.29}^{+0.35}$ & $1.55_{-0.27}^{+0.33}$ & $1.48_{-0.22}^{+0.26}$ & $1.51_{-0.22}^{+0.26}$ \\[1ex]
           $c$ & $10.1_{ -5.4}^{+11.6}$ & $ 8.1_{ -2.1}^{ +2.8}$ & $10.2_{ -3.3}^{ +4.8}$ & $ 8.7_{ -1.9}^{ +2.4}$ \\[0.5ex]
\cmidrule(lr){1-5}
          $\lg\,M^\text{a}$ & $ 0.19 \pm 0.09$ & $ 0.19 \pm 0.08$ & $ 0.17 \pm 0.07$ & $ 0.18 \pm 0.07$ \\
          $\lg\,c$ & $ 1.01 \pm 0.33$ & $ 0.91 \pm 0.13$ & $ 1.01 \pm 0.17$ & $ 0.94 \pm 0.10$ \\
$\rho_\text{corr}$ & $          0.24$ & $          0.10$ & $         -0.42$ & $         -0.26$ \\
\midrule
\multicolumn{3}{l}{EAGLE}\\
\cmidrule(lr){1-5}
  $M^\text{a}$ & $1.29_{-0.20}^{+0.24}$ & $\bm{1.23_{-0.18}^{+0.21}}$ & $1.27_{-0.15}^{+0.17}$ & $1.26_{-0.15}^{+0.17}$ \\[1ex]
           $c$ & $11.0_{ -3.3}^{ +4.8}$ & $\bm{ 9.4_{ -2.1}^{ +2.8}}$ & $11.7_{ -2.5}^{ +3.2}$ & $10.4_{ -1.9}^{ +2.3}$ \\[0.5ex]
\cmidrule(lr){1-5}
          $\lg\,M^\text{a}$ & $ 0.11 \pm 0.07$ & $\bm{ 0.09 \pm 0.07}$ & $ 0.10 \pm 0.06$ & $ 0.10 \pm 0.06$ \\
          $\lg\,c$ & $ 1.04 \pm 0.16$ & $\bm{ 0.97 \pm 0.11}$ & $ 1.07 \pm 0.10$ & $ 1.02 \pm 0.09$ \\
$\rho_\text{corr}$ & $          0.48$ & $\bm{          0.34}$ & $          0.06$ & $          0.05$ \\
\midrule \midrule 
\multicolumn{5}{l}{$^\text{a}$ In units of $10^{12}\msun$ and including the baryonic contribution.}
\end{tabular}
\label{tab:esti_result}
\end{table}

As our DF model is constructed to be the average DF for a sample of halos under a set of assumptions,
individual halos are expected to deviate from the model in several aspects. 
For example, the mass distribution may deviate from a perfect spherical NFW profile, 
the satellites may be neither fully phase-mixed nor mutually independent due to the hierarchical accretion,
and the scaled DF may not exactly follow our proposed form.
The presence of massive satellites or companion galaxies, e.g., the LMC or M31 for our MW, might further increase the deviations (see more discussion below).
All of these deviations can contribute to the halo-to-halo scatter in our mass estimates besides the statistical uncertainty.
Using a large mock sample of realistic halos from the SAM-MII simulation,
\citet{Li2019} estimated a systematic uncertainty of $\sim 7\%$ (0.03 dex) in $M$
when the prior based on the $M$-$c$ relation was used.
It is worth emphasizing that given a realistic mock sample, all of the above halo-to-halo scatters should have already been captured by the above uncertainty.
As discussed in \refsec{sec:hydro_dep}, the dependence of our DF
on the hydrodynamics-based simulations introduces an additional systematic uncertainty
of $\sim 5\%$ in $M$. However, the above systematic uncertainties in $M$ are significantly 
smaller than the current statistical uncertainty of $\sim 17\%$
(see the relevant bold entry in \reftab{tab:esti_result}).

\subsection{Robustness of results} \label{sec:robust}

We now demonstrate the robustness of our results.
For the EAGLE DF (see Appendix~\ref{sec:massive_test}), we find that
the massive neighbor M31 has no significant influence on the inferred MW halo properties.
The influence of the LMC might be more complicated.
This massive satellite could imply a particular assembly history of the MW
and induce non-trivial reflex motion of other satellites and the MW stellar halo (e.g., \citealt{Petersen2020,Erkal2020}),
thereby possibly causing bias in the MW mass estimate.
Using a simplified test, \citet{Li2017} showed that
adding a velocity offset of $30\,\kms$ to the MW
to mimic the reflex motion caused by the LMC
only changes the results at a level $\lesssim 3\%$.
Because the reflex motion is more complex than a simple bulk motion,
it is better captured by our adopted simulations, which automatically include the effects due to massive satellites.
As shown by the tests based on the simulations in Appendix \ref{sec:massive_test},
the LMC has no significant influence on our MW mass estimate.
Nevertheless, it is worth quantifying the effects of the LMC more precisely with a larger halo sample in the future.
Below we present more tests and focus on the EAGLE DF
with the prior based on the $M$-$c$ relation. The same tests for the SAM-MII DF give similar
conclusions.

\begin{figure}[htbp]
  \centering
  \includegraphics[width=0.45\textwidth]{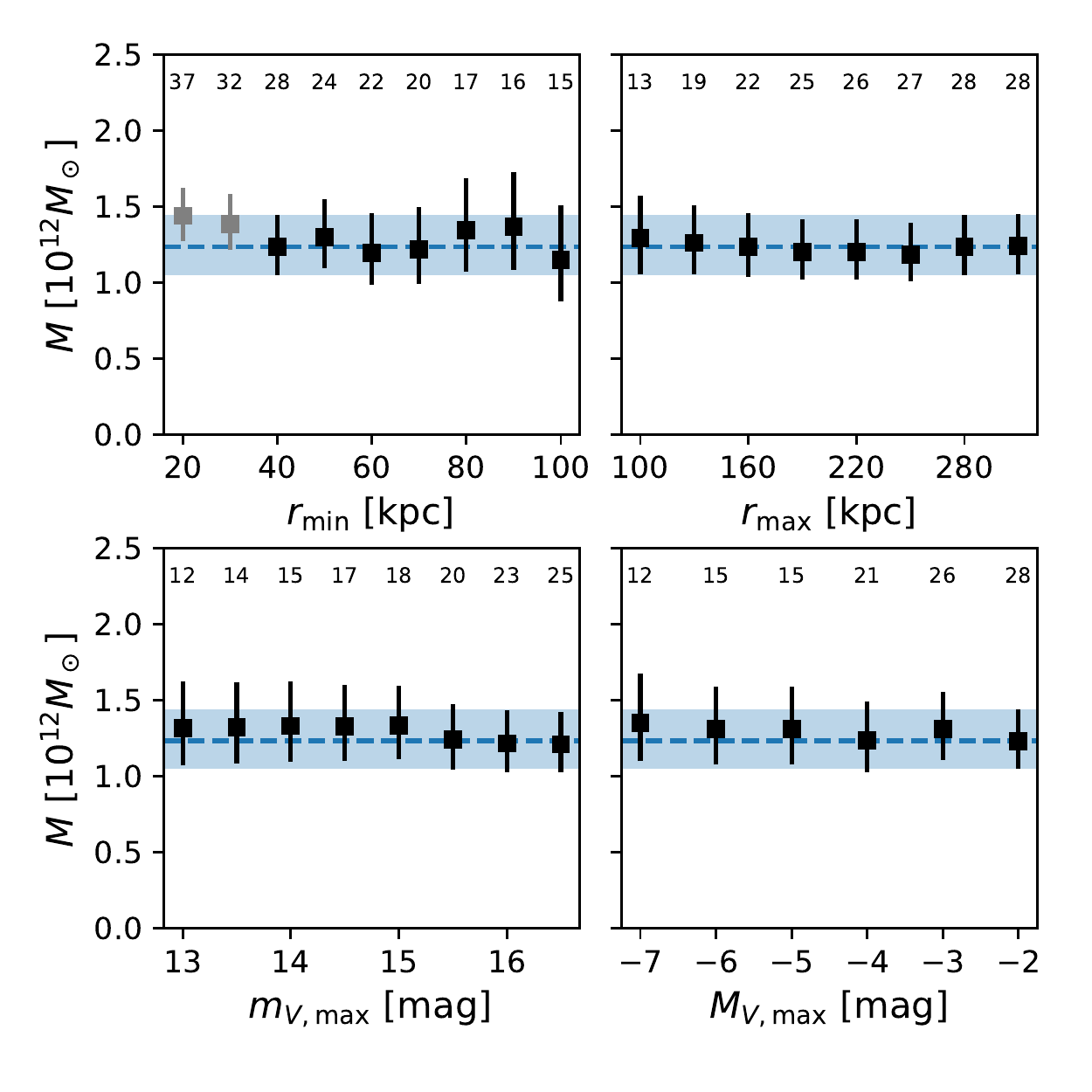}
  \vspace{-4mm}
  \caption{
  Robustness tests for the MW halo mass $M$ inferred from the EAGLE DF.
  In each panel, $r_{\min}$, $r_{\max}$,
  maximum apparent magnitude $m_{V,{\rm max}}$,
  or maximum absolute magnitude $M_{V,{\rm max}}$ for the satellite sample is 
  changed from the fiducial value.
  For each sample, a filled square with an error bar shows the inferred $M$ with 
  the $1\sigma$ uncertainty. The number of satellites, $n_\mathrm{sat}$, in each
  sample is indicated above the corresponding square. For reference,
  results (gray color) are also shown for two samples including satellites within $40\,\kpc$, 
  which are beyond the scope intended for our method.
  The dashed line and the shaded band show the results for the fiducial sample 
  ($n_\mathrm{sat}=28$, see \reffig{fig:obs_selection}).
  }
  \label{fig:test_robust}
\end{figure}

From a jackknife (leave-one-out) test on our sample of satellites, we find that the scatter in the inferred 
MW halo mass $M$ is comparable to the estimated statistical uncertainty in \reftab{tab:esti_result}.
In addition, the effect on the inferred $M$ is negligible compared to the statistical uncertainty
when we exclude from our sample all of the possible LMC satellites: the Small Magellanic Cloud, 
Fornax, Carina I, and Horologium I \citep{Kallivayalil2018,Pardy2019}.
Finally, as shown in \reffig{fig:test_robust}, we get remarkably consistent results on $M$
when varying the sample selection criteria based on the distance interval ($r_{\min},\ r_{\max}$), 
brightness ($m_{V,{\rm max}}$), or luminosity ($M_{V,{\rm max}}$) of the satellites. 
Note that varying $M_{V,\max}$ only changes the satellite sample, 
but the analysis remains the same as for the fiducial case.
For the other tests, the $r_{\min}$, $r_{\max}$, and $\Robsmax$
used in \refeqn{eqn:selection} are changed accordingly.
In particular, $\Robsmax$ depends on $m_{V,{\rm max}}$.

The robustness of our results demonstrated by the above tests can be attributed to two factors.
First, the constraining power mainly comes from the bright satellites with precise measurements.
Therefore, so long as a sample includes a sufficient number of such satellites,
the inferred $M$ and its uncertainty should not change very much as the sample varies.
More importantly, the robustness of our results also reflects the validity of our method, 
especially in treating the selection function and observational errors.
Ignoring observational uncertainties overestimates the halo mass ($M\simeq 1.8\times 10^{12}\msun$)
and gives an absurdly large concentration ($c\simeq 40$), while ignoring the selection function
severely overestimates the concentration ($c \simeq 17$).

We emphasize that a rigorous and straightforward treatment of the selection function and 
observational errors is an important feature of the DF method. In contrast, it is rather
difficult to treat observational errors in methods based on the Jeans equation.
In some previous studies using such methods, because observational errors were not treated properly,
including Leo I or not can change the estimated MW halo mass by $\sim 30\%$ \citep[e.g.,][]{Watkins2010a}.

Another possible concern is the flattened satellite distribution of the MW,
though its cosmological significance is still under debate
(e.g., \citealt{Pawlowski2013,Cautun2015a,Shao2019}).
However, the anisotropic distribution of satellites 
is unlikely able to bias our result significantly for two reasons.
First, the mass estimate relies on the distance and velocity 
rather than the orbital orientation of a satellite.
Second, as shown by the extensive tests in \citet{Li2019} and Appendix~\ref{sec:massive_test},
our method is robust for halos of a very wide range of halo structure, formation history and environment.
Nevertheless, 
it is worth further investigating the peculiarities of the MW and their potential influence on the mass estimate.

\subsection{Comparison of DFs with observations}\label{sec:model_selec}

For direct validation of the SAM-MII and EAGLE DFs, as well as the associated estimates of the MW halo properties,
we compare these DFs with the observed satellite kinematics.
For this purpose, we use the best-fit $M$ and $c$ inferred from each DF with the prior based on the $M$-$c$ relation.
Under our assumptions, the DF $f({\bm{r}},{\bm{v}})$ 
can be written as $f(E,L)$, where $E$ and $L$ are functions of $r$, the radial velocity $\vr$,
and the tangential velocity $\vt$ (see \refsec{sec:method-DF}).
Because it is difficult to show $f({\bm{r}},{\bm{v}})$ in the 3D space of $r$, $\vr$, and $\vt$,
we instead display the projected DF in the 2D space of $r$ and $\vt$ by marginalizing $\vr$ and
taking into account the selection function
\begin{equation}
    p_\mathrm{s} (r, \vt) \propto 
        8 \pi^2 r^2 \vt N (< M_{V\!,\,\mathrm{lim}} (r))\int f ({\bm{r}}, {\bm{v}}) d\vr,
    \label{eqn:obs_kin}
\end{equation}
where the factor $8 \pi^2 r^2 \vt$ comes from the differential phase-space volume element,
$N(<M_V) = 10^{0.156 M_V +2.21}$ is the complete satellite luminosity function
derived by \citet{Newton2018}, and $M_{V\!,\,\mathrm{lim}}(r)$ is the limiting absolute magnitude for 
{\it Gaia} proper motion measurement at radius $r$ [i.e., $r=\Robsmax(M_{V\!,\,\mathrm{lim}}(r))$, see
\refsec{sec:selec_func}].

\begin{figure*}[htb!]
  \centering
  \includegraphics[width=0.45\textwidth]{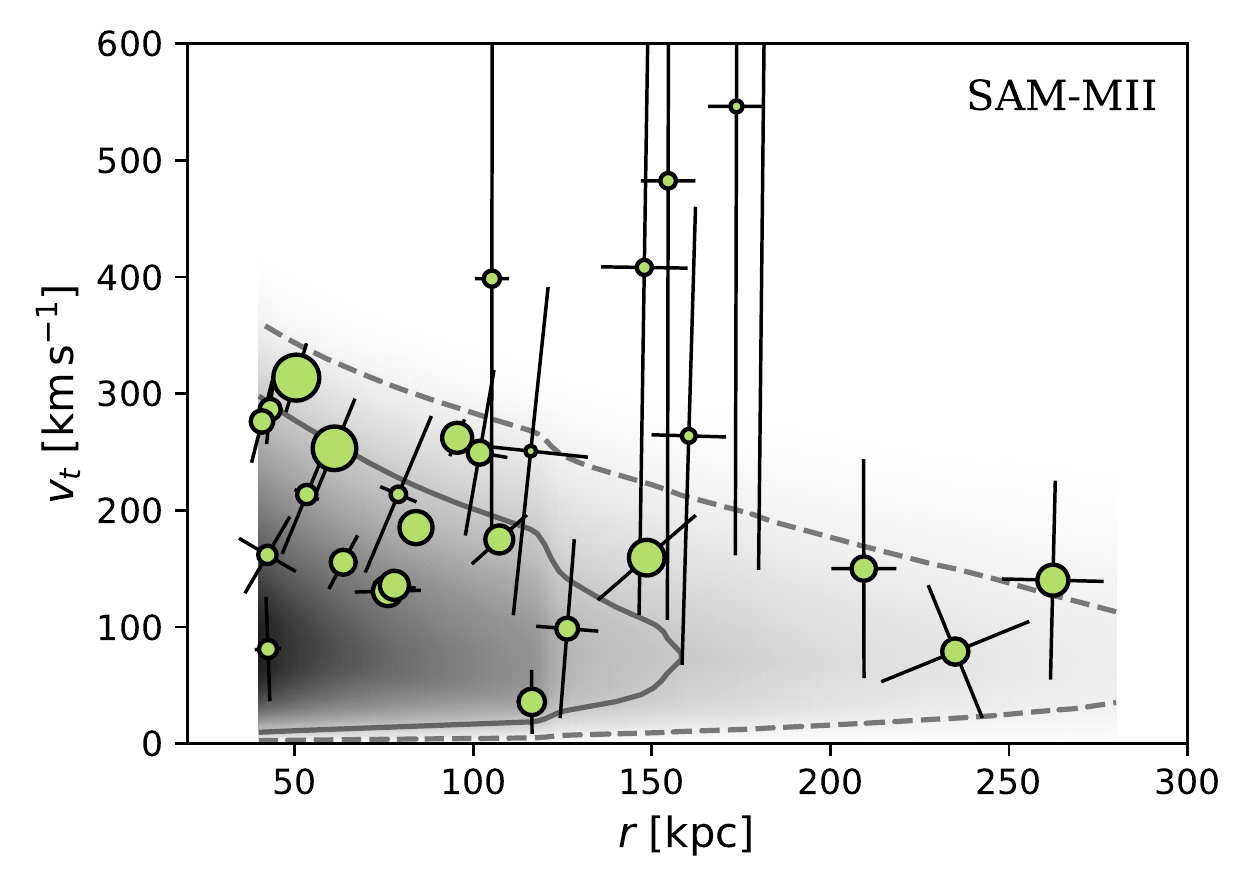}
  \includegraphics[width=0.45\textwidth]{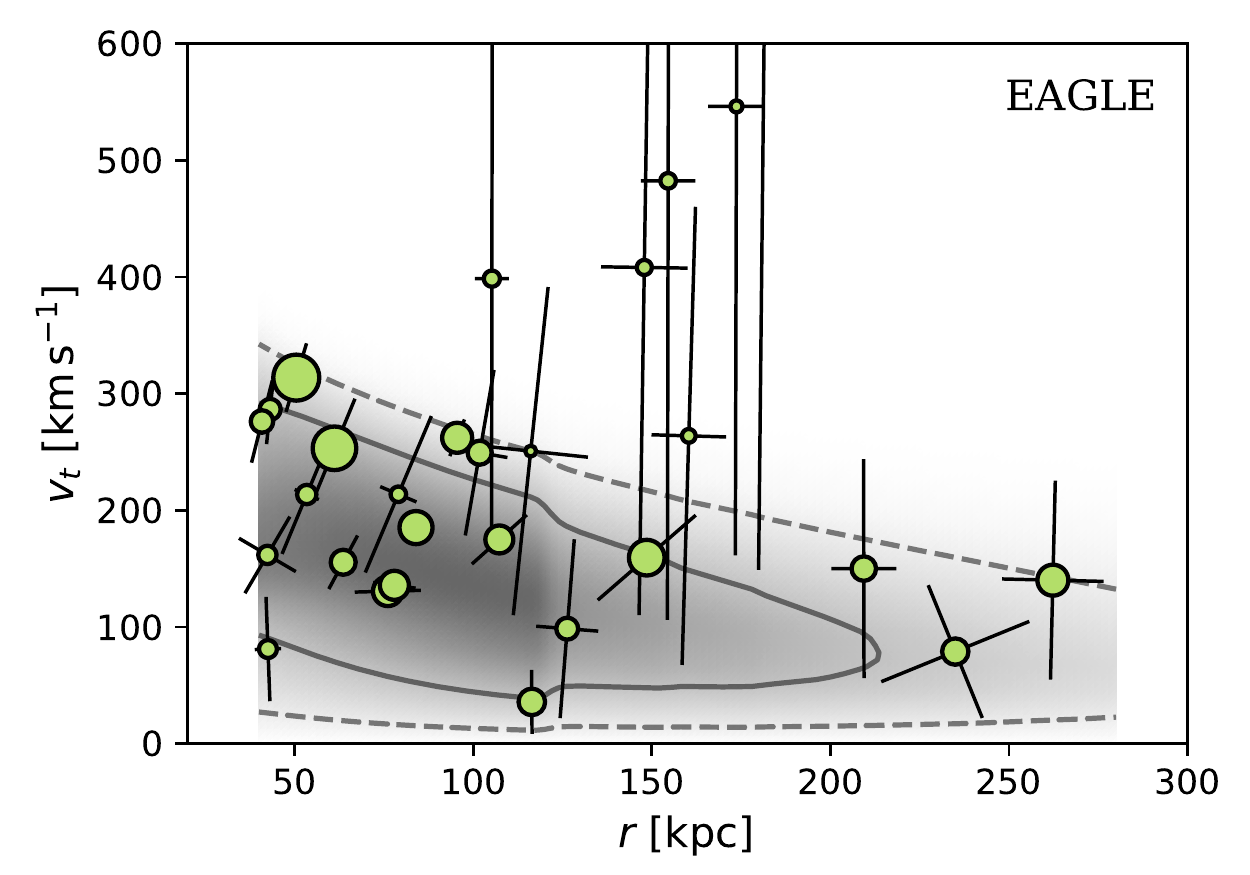}
  \caption{
  Comparison of the projected SAM-MII (left panel) and EAGLE (right panel) DFs with the observations.
  In each panel, $p_\mathrm{s} (r, \vt)$ with the best-fit $M$ and $c$ is shown as
  shades of gray (on the same intensity scale) along with the $1\sigma$ (solid) and $2\sigma$ (dashed) 
  confidence contours. The data for our satellite sample are shown as circles, with larger ones
  indicating satellites with higher brightness. The error bars on each circle represent the axes of the $1\sigma$ 
  error ellipse. See text for details.
  }
  \label{fig:obs_kin2}
\end{figure*}

\reffig{fig:obs_kin2} shows $p_\mathrm{s} (r, \vt)$ for satellites with $40<r<280\,\kpc$
as shades of gray along with $1\sigma$ and $2\sigma$ confidence contours.\footnote{
The abrupt changes at $r\sim 120$ kpc in the confidence contours are caused by the selection function.
See footnote 2 for details.}
Because the observational errors vary greatly among satellites, we have not included them
in deriving $p_\mathrm{s} (r, \vt)$ for simplicity. Instead, we include these errors\footnote{ 
The error bars on a data symbol represent the axes of
the $1 \sigma$ error ellipse, which is determined from the Monte Carlo realizations of the data
(see \refsec{sec:satellites}) using the minimum covariance determinant method \citep{Hubert2018}.}
when showing the kinematic data for our satellite sample in \reffig{fig:obs_kin2}. 
Taking these errors into account when comparing the distribution of the data points 
with respect to the shades of gray and the confidence contours, we find that 
for both the SAM-MII and EAGLE DFs, the $2\sigma$ confidence region of $p_\mathrm{s} (r, \vt)$ 
with the best-fit $M$ and $c$ is consistent with the observations.

\reffig{fig:obs_kin2} also shows that the EAGLE DF provides a significantly better match to 
the observations than the SAM-MII DF. Specifically, compared with the observations, the SAM-MII DF 
predicts a distribution of satellites that is too concentrated at smaller $r$ and $\vt$.
This discrepancy was also noticed
for a similar halo sample based on the SAM-MII \citep{Cautun2016} or APOSTLE simulation suite \citep{Riley2018}.
The above results are consistent with the ratio of the Bayesian evidence $Z$ 
[see Equation (\ref{eqn:posterior})] for the two DFs, which is also known as the Bayes factor.
We find $Z(\mathrm{EAGLE})/Z(\mbox{SAM-MII}) = 33$
(or 25 when the flat prior on $\lg c$ is used), which indicates that the observations
strongly favor the EAGLE DF. Therefore, the results from the EAGLE DF are recommended.

\subsection{Inferring MW satellite kinematics} \label{sec:reduced_kin}

The orbits of satellites can shed important light on their past evolution and the assembly history of the MW.
However, as shown in \reffig{fig:obs_kin2}, distant satellites typically have poorly measured proper motion, 
which makes it difficult to calculate their precise orbits. Having shown that the EAGLE DF provides a good
description of the MW satellite kinematics, we can now use it to infer more precise velocities for those
satellites with poor current measurements.

Given the kinematic data $\{\hat{\bm{w}}_j\}_{j=1,\ldots,n_\mathrm{sat}}$ for $n_\mathrm{sat}$ satellites, 
the posterior distribution of the true kinematics for the $i$th satellite is
\begin{equation}
\begin{split}
  p (\bm{w}_i| \{\hat{\bm{w}}_j\}) \propto 
        \int & p_{\mathrm{err}} (\hat{\bm{w}}_i | \bm{w}_i ) p_\mathrm{s} (\bm{w}_i | M, c)\\
      \times & p (M, c| \{\hat{\bm{w}}_{j \neq i}\}) d M d c,
\label{eqn:kinpost}
\end{split}
\end{equation}
where $p (M, c| \{\hat{\bm{w}}_{j \neq i}\})$ is the distribution of halo parameters
inferred from the data on all of the other satellites [see \refeqn{eqn:posterior}].
We calculate $p (\bm{w}_i| \{\hat{\bm{w}}_j\})$ using \textit{importance sampling}.
We first generate Monte Carlo realizations $\{\bm{w}_{ik} \}_{k=1,\,2,\cdots}$ with
$p (\bm{w}_{ik}) \propto p_{\mathrm{err}} (\hat{\bm{w}}_i | \bm{w}_{ik})$ (see \refsec{sec:satellites}).
These $\bm{w}_{ik}$ along with the corresponding importance weight $\int p_\mathrm{s} (\bm{w}_{ik} | M, c) p (M, c| \{\hat{\bm{w}}_{j \neq i}\}) d M d c$
represent the weighted realizations of $p (\bm{w}_i| \{\hat{\bm{w}}_j\})$,
from which we can infer the best-fit values of $\bm{w}_i$ and the associated uncertainties.

\begin{figure}[htbp]
  \centering
  \includegraphics[width=0.45\textwidth]{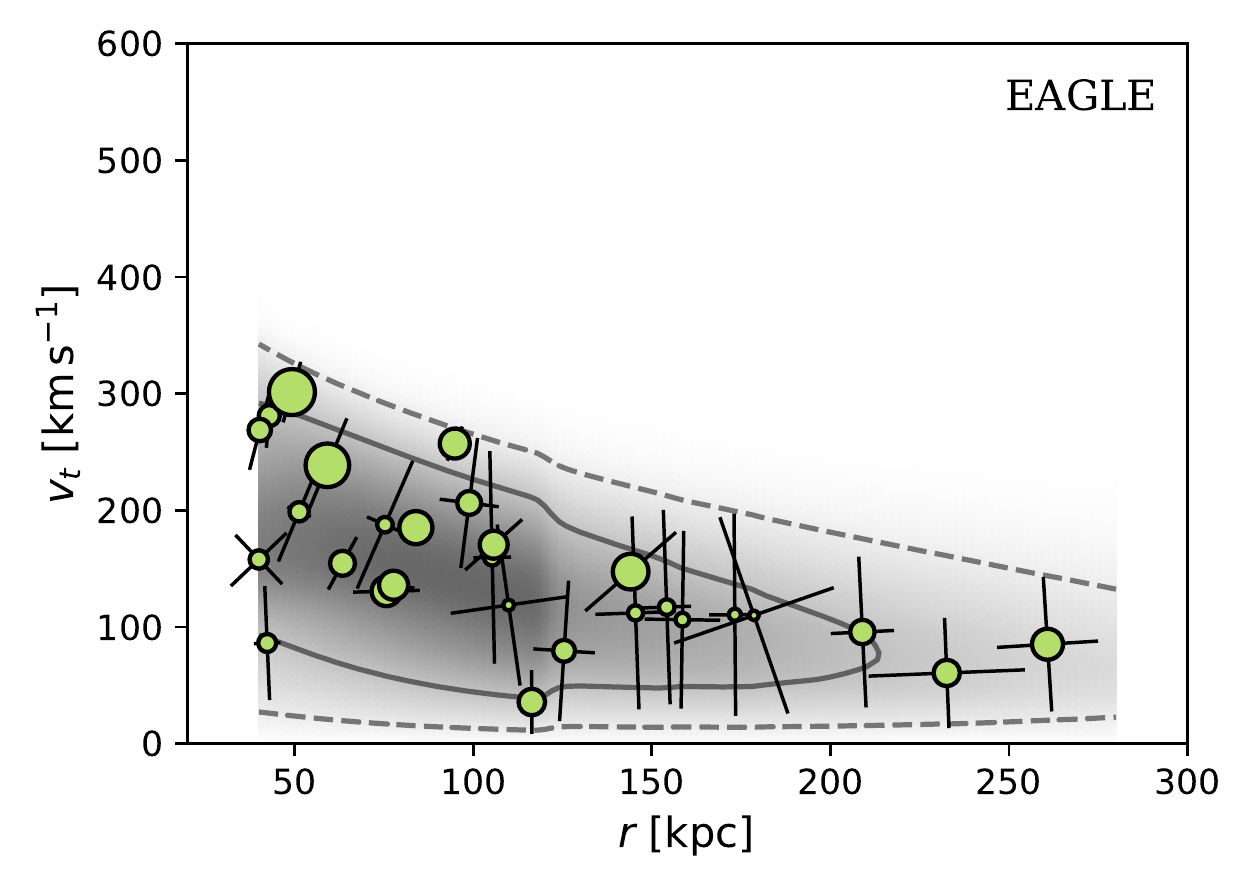}
  \caption{
  Same as the right panel of \reffig{fig:obs_kin2}, but showing the posterior satellite kinematics.
  }
  \label{fig:reduced_kin}
\end{figure}

The posterior satellite kinematic data inferred from \refeqn{eqn:kinpost}
are shown in \reffig{fig:reduced_kin}.
It can be seen that the uncertainties in $\vt$ are greatly reduced for those satellites with poor
current measurements. As expected, the overall distribution of the posterior satellite kinematics
also becomes very close to the projected DF $p_\mathrm{s} (r, \vt)$ (see \refsec{sec:model_selec}).
The posterior kinematic data are given in Appendix \ref{sec:data_table},
are available online at \url{https://github.com/syrte/mw_sats_kin},
and are archived in China-VO (doi:\href{https://doi.org/10.12149/101018}{10.12149/101018}).

\subsection{Dependence of the DF on cosmological simulations} \label{sec:hydro_dep}

\reftab{tab:esti_result} shows that the best-fit MW halo mass from the SAM-MII DF with the prior based on the $M$-$c$ relation
is $\sim 26\%$ larger than that from the EAGLE DF. In addition,
tests with mock samples of EAGLE halos show that the SAM-MII DF overestimates the halo mass by $\sim 12\%$ on average.
Below, we discuss the underlying cause for the difference between these two DFs, which in turn gives rise to different estimates
of halo properties.

The SAM-MII and EAGLE simulations differ in that the latter is based on hydrodynamics 
with baryonic physics.
We consider that the full treatment of the stellar disk, including its gravitational effects, 
by the EAGLE simulation is most likely the main cause for the difference between
the SAM-MII and EAGLE DFs.\footnote{
    It is well known that the density profile contracts in hydrodynamics-based simulations
    and the satellite kinematics responds accordingly.
    However, the dimensionless DF should remain similar so long as the NFW profile still applies to the outer halo.
    Therefore, the contraction of the density profile is unlikely the main 
    cause for the difference between the SAM-MII and EAGLE DFs.}
The stellar disk enhances the tidal field in the inner halo, thereby increasing
the disruption rate for satellites with small pericenter distances $r_\mathrm{peri}$
\citep[e.g.,][]{Garrison-Kimmel2017,Sawala2017,Richings2018}.
Because the formation and growth of the stellar disk were treated in the SAM-MII simulation 
without accounting for the associated change in the gravitational field,
more satellites with small $r_\mathrm{peri}$ survived in this simulation compared to the EAGLE simulation 
and the MW observations. Consequently, satellites with small $r$ and $\vt$, which also have small $r_\mathrm{peri}$, are over-represented by
the SAM-MII DF (see \reffig{fig:obs_kin2}).

We find that the radial phase angle is uniformly distributed
on average for satellites in both the SAM-MII \citep{Li2019} and EAGLE simulations.
So enhanced disruption by the stellar disk is more of a selection on orbit than on phase angle.
Guided by this result, we mimic the gravitational effects of the stellar disk by manually increasing 
the disruption rate for satellites on orbits with small $r_\mathrm{peri}$ in the SAM-MII simulation.
As shown in \reffig{fig:destruction}, this prescription (see Appendix~\ref{sec:hydro_esti})
can give a projected DF $p_\mathrm{s} (r, \vt)$ very similar to that for the EAGLE simulation.
Compared with $M=1.55_{-0.27}^{+0.33}\times 10^{12}\,\msun$ from the SAM-MII DF, the estimate from this 
modified SAM-MII DF, $M=1.35_{-0.19}^{+0.23}\times 10^{12}\,\msun$, is also much closer to $M=1.23_{-0.18}^{+0.21}\times 10^{12}\,\msun$ 
from the EAGLE DF.

While the hydrodynamics-based EAGLE simulation matches the observations better than the SAM-MII simulation,
variation in the treatment of physical processes in current hydrodynamics-based simulations also leads to
scatter in estimate of halo properties from the DF method.
For example, compared to the APOSTLE simulation, the central galaxies in the Auriga simulation are more 
massive and, hence, more efficient at disrupting satellites. Consequently, the latter has approximately three times
fewer surviving satellites within $0.1 R$ than the former \citep{Richings2018}.
Similar to the comparison of the EAGLE and SAM-MII DFs, the Auriga DF is expected to give lower halo mass 
estimates than the APOSTLE DF. This scatter in the halo mass estimate for DFs from hydrodynamics-based simulations 
should be much smaller than the difference of $\sim 12\%$ for the SAM-MII and EAGLE DFs.
To better quantify this uncertainty, we vary the enhanced satellite disruption in the SAM-MII simulation
according to the prescription in Appendix~\ref{sec:hydro_esti}, and obtain new satellite samples to construct
modified SAM-MII DFs. Applying these DFs to EAGLE halos shows a scatter of only $\sim 5\%$ in the halo mass estimate
(see Appendix~\ref{sec:hydro_esti}).
We take this result as a reasonable estimate of the scatter for DFs from hydrodynamics-based simulations.
This estimate is consistent with the findings of \citet{Callingham2018}, who recovered halo masses in the
Auriga simulation with little bias using the orbital distribution from the EAGLE simulation.

\begin{figure}[tbp]
  \centering
  \includegraphics[width=0.45\textwidth]{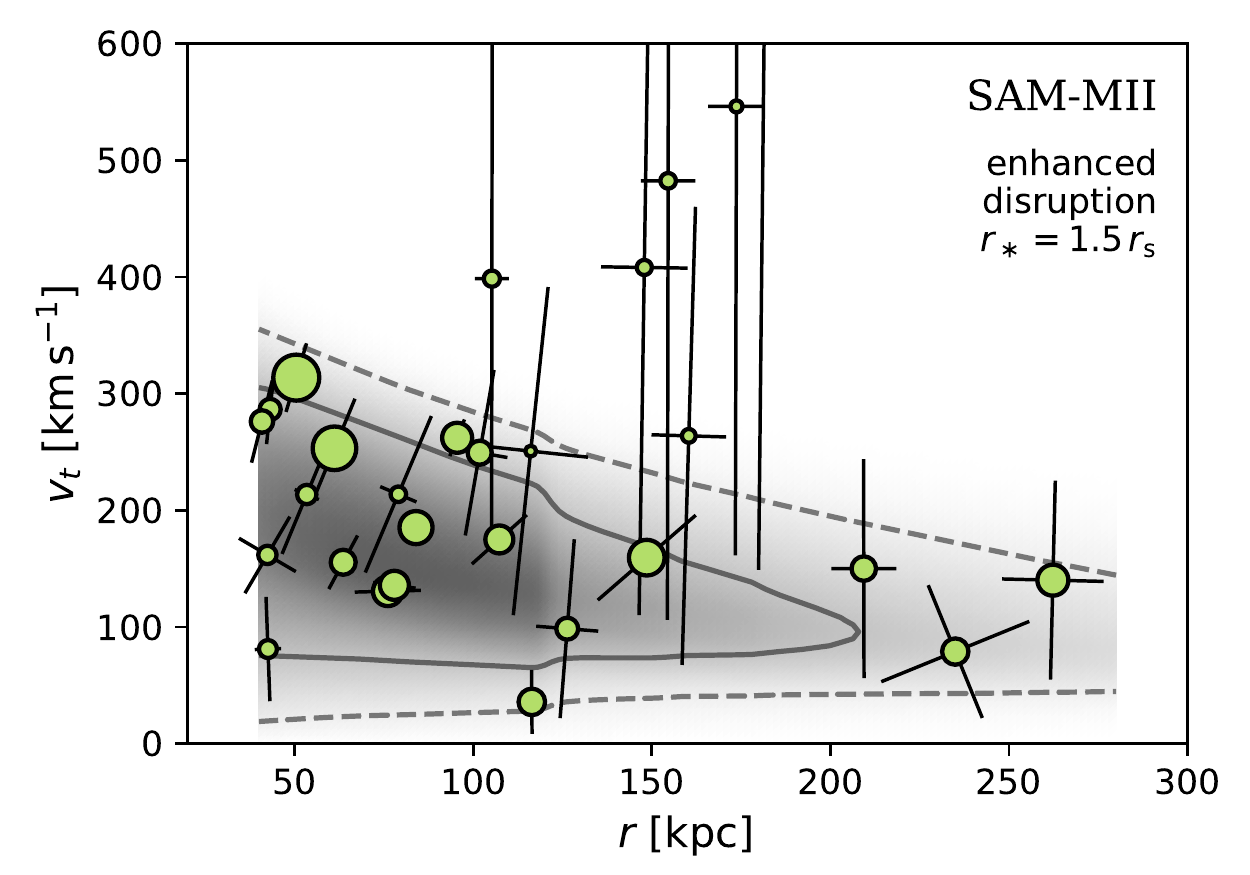}
  \caption{
  Same as the left panel of \reffig{fig:obs_kin2}, 
  but for the modified SAM-MII DF based on enhanced disruption of satellites with small pericenter distances. 
  See Appendix~\ref{sec:hydro_esti} for prescription of enhanced disruption
  using $r_\ast$.
  }
  \label{fig:destruction}
\end{figure}

\section{Comparison with previous works and joint constraints} \label{sec:comp_joints}

In this section, we compare the MW mass and its distribution inferred from the EAGLE DF
with results from previous works. We also discuss possible improvement of our results by
combining different tracer populations.

\subsection{Comparison with previous results} \label{sec:compare}

\begin{figure*}[htb]
  \centering
  \includegraphics[width=0.45\textwidth]{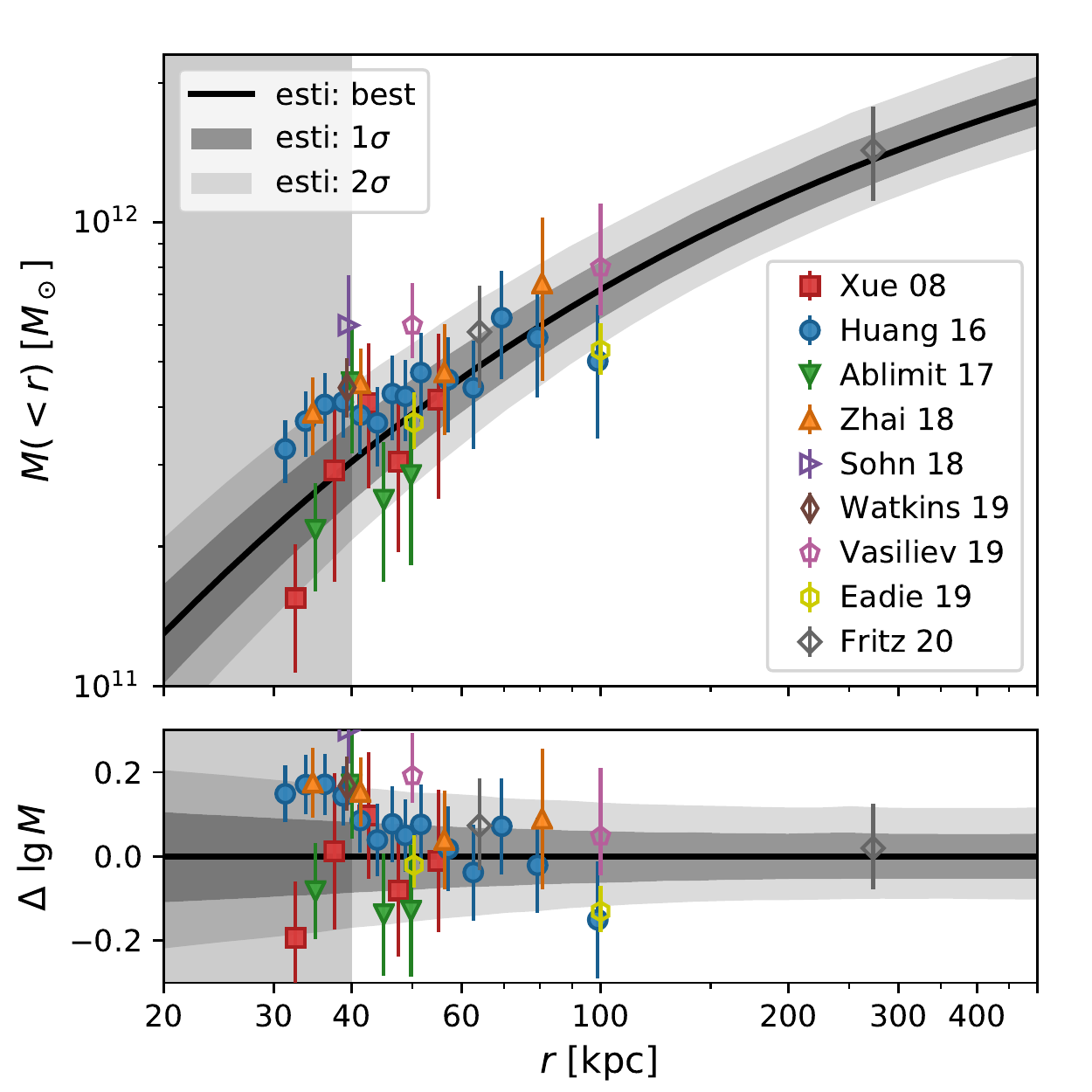}
  \includegraphics[width=0.45\textwidth]{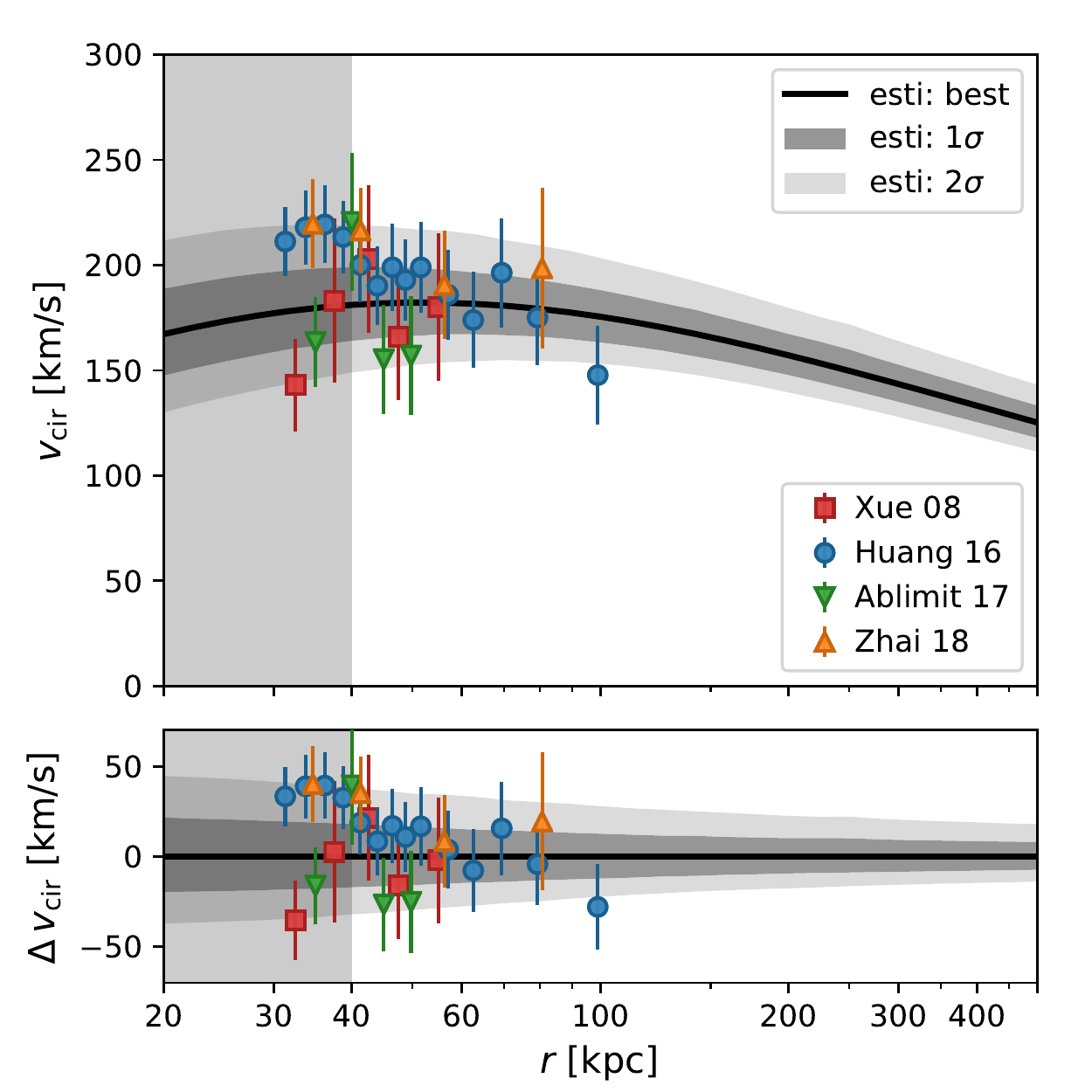}
  \caption{
  Comparison of the inferred mass profile (left panel) and rotation curve (right panel) 
  for the MW outer halo with previous measurements (symbols with error bars).
  The black curves show the best-fit results from the EAGLE DF with the prior based on the $M$-$c$ relation,
  and the associated shaded bands are the $1\sigma$ and $2\sigma$ confidence regions.
  While our method focuses on the outer halo with $r>40\,\kpc$, our results 
  should still be reliable inside but not too far from $r=40\,\kpc$.
  }
  \label{fig:compare}
\end{figure*}

Many studies were dedicated to measuring the halo mass and its distribution for the MW
(for a comprehensive review, see \citealt{Wang2019}). In particular,
much work focused on the rotation curve (RC) or masses enclosed within certain radii.
A selected collection of recent measurements
with halo stars \citep{Xue2008,Huang2016, Ablimit2017, Zhai2018}, globular clusters 
\citep{Sohn2018,Watkins2018,Vasiliev2018,Eadie2018},
and satellites \citep{Fritz2020} beyond $r=40\,\kpc$
is shown in \reffig{fig:compare}. We convert the RC into the mass profile using 
$M(<r)=r v_\mathrm{cir}^2/G$ and vice versa. Here $M(<r)$ is the mass enclosed within 
radius $r$ and $v_\mathrm{cir}$ is the circular velocity at this $r$.
\reffig{fig:compare} also shows our results inferred from the EAGLE DF with 
the prior based on the $M$-$c$ relation for comparison
(see \citealt{Eadie2018} and \citealt{Wang2019} 
for a more comprehensive comparison).

It can be seen from \reffig{fig:compare} that our results are in good agreement with
the RC measurements (within $1\sigma$ for most cases).
Note that when multiple models of velocity anisotropy $\beta$ were used for 
an RC dataset, only those results assuming relatively high $\beta$ are shown based on the
recent measurement of $\beta$ for halo stars with proper motion from {\it Gaia} \citep{Bird2018}.
The low $\beta$ found in some earlier studies is likely due to e.g., contamination from the disk 
\citep{McMillan2017} and substructures \citep{Loebman2018}.

We also note that studies using halo stars typically favor a smaller MW halo mass
(e.g., $M\simeq 0.8\times 10^{12}\msun$ from \citealt{Xue2008,Huang2016})
and a higher concentration \citep[$c\sim 14$--20, e.g.,][]{Deason2012,Kafle2014,Huang2016,Zhai2018}.
These differences from our results are likely due to the profile extrapolation to the outer halo used
in these studies. For example, ignoring the contraction of dark matter profile in the inner halo
would lead to biased profile extrapolation \citep{Cautun2019}. Because we use satellites, which
are the proper tracers of the outer halo, the above issue is irrelevant for our results.
Remarkably, our inferred mass profile is in very good agreement (within $\sim 0.5 \sigma$)
with the corresponding result of \citet{Cautun2019}, who used both halo stars and satellites as tracers,
and with that of \citet{Fritz2020}, who applied the mass estimator of \citet{Watkins2010a} to satellites 
within multiple radii.

\subsection{Joint constraint with RC from halo stars} \label{sec:joints}

Combining different tracer populations on different spatial scales 
can improve the constraint on the MW mass profile.
While the halo mass is mainly constrained by distant tracers like satellites,
the nearby tracers serve as a better probe of the inner profile and, therefore, 
can improve the estimate of the halo concentration.
In addition, if different tracer populations have independent systematics,
combining them can reduce the systematic uncertainties.
Examples of combining different tracer populations to constrain the MW halo properties include 
\citet{McMillan2011,McMillan2017} and \citet{Nesti2013} for using gas clouds, masers, and stars and 
\citet{Callingham2018} for using satellites and globular clusters.

For illustration, here we combine satellite kinematics with the RC from halo stars 
to constrain the MW halo mass and concentration. Using $\sim\! 5,700$ halo K giants
selected from the SDSS/SEGUE survey, \citet{Huang2016} derived the RC for the outer halo based on
the spherical Jeans equation.
While their data could benefit from a reanalysis using an updated $\beta$ from {\it Gaia},
these data are currently the best for relatively large radii.
We only use their data for $40<r<80\,\kpc$ (see the right panel of \reffig{fig:compare}).
An important issue is the treatment of the relevant uncertainties.
In addition to the measurement uncertainty $\sigma_{v_{\mathrm{cir}, i}}$ in the
circular velocity $v_{\mathrm{cir},i}$ at radius $r_i$, there is an additional large 
systematic uncertainty from the assumed power-law index $\alpha$ for the stellar density 
profile in the outer halo. \citet{Huang2016} adopted $\alpha=-4.5$ as the fiducial value.
However, current observations allow $\alpha=-3.8$ to $-5$ and variation over this range
systematically changes the derived $v_{\mathrm{cir},i}$ at the level of 
$\sigma_\mathrm{sys}= 15\,\kms$ (see \citealt{Huang2016} for detailed discussion).
Therefore, the RC measurements at different radii are not independent. Ignoring 
this correlation of measurements, as usually done in previous studies, leads to 
underestimated formal errors. A proper treatment is to use the covariance matrix
\begin{equation}
    \mathcal{M}_{i j} = \mathrm{cov}\!\left(v_{\mathrm{cir}, i},\, v_{\mathrm{cir}, j}\right) =  \Bigg\{
     \begin{aligned}
        &\sigma_{v_{\mathrm{cir}, i}}^2 + \sigma_\mathrm{sys}^2, & i=j,\\
        &\sigma_\mathrm{sys}^2, & i\neq j.
     \end{aligned}
\end{equation}

The RC data can be modeled as a multivariate Gaussian distribution.
For a specific set of $M$ and $c$ for the NFW profile, the expected $v_{\mathrm{cir}}$ at radius $r_i$ is
$v_{\mathrm{cir}} (r_i |M, c)=\sqrt{G M_\mathrm{NFW}(<r_i|M, c)/r_i}$.
The probability (likelihood) of $n$ measurements $\{ v_{\mathrm{cir},i}\}_{i=1,2,\ldots,n}$ is
\begin{equation}
    p (\left\{ v_{\mathrm{cir},i} \right\} |M, c) 
    = \frac{\exp\! \big[ -\! \frac{1}{2} \sum_{i, j} \Delta v_i (\mathcal{M}^{- 1})_{i j} \Delta v_j \big]}
           {\sqrt{(2 \pi)^n \det (\mathcal{M})}},
\label{eqn:RC}
\end{equation}
where $\Delta v_i = v_{\mathrm{cir}, i} - v_{\mathrm{cir}} (r_i |M, c)$.
The above likelihood can be used independently, or multiplied by the likelihood in \refeqn{eqn:posterior}
for joint analysis.

\begin{figure}[tbp]
  \centering
  \includegraphics[width=0.45\textwidth]{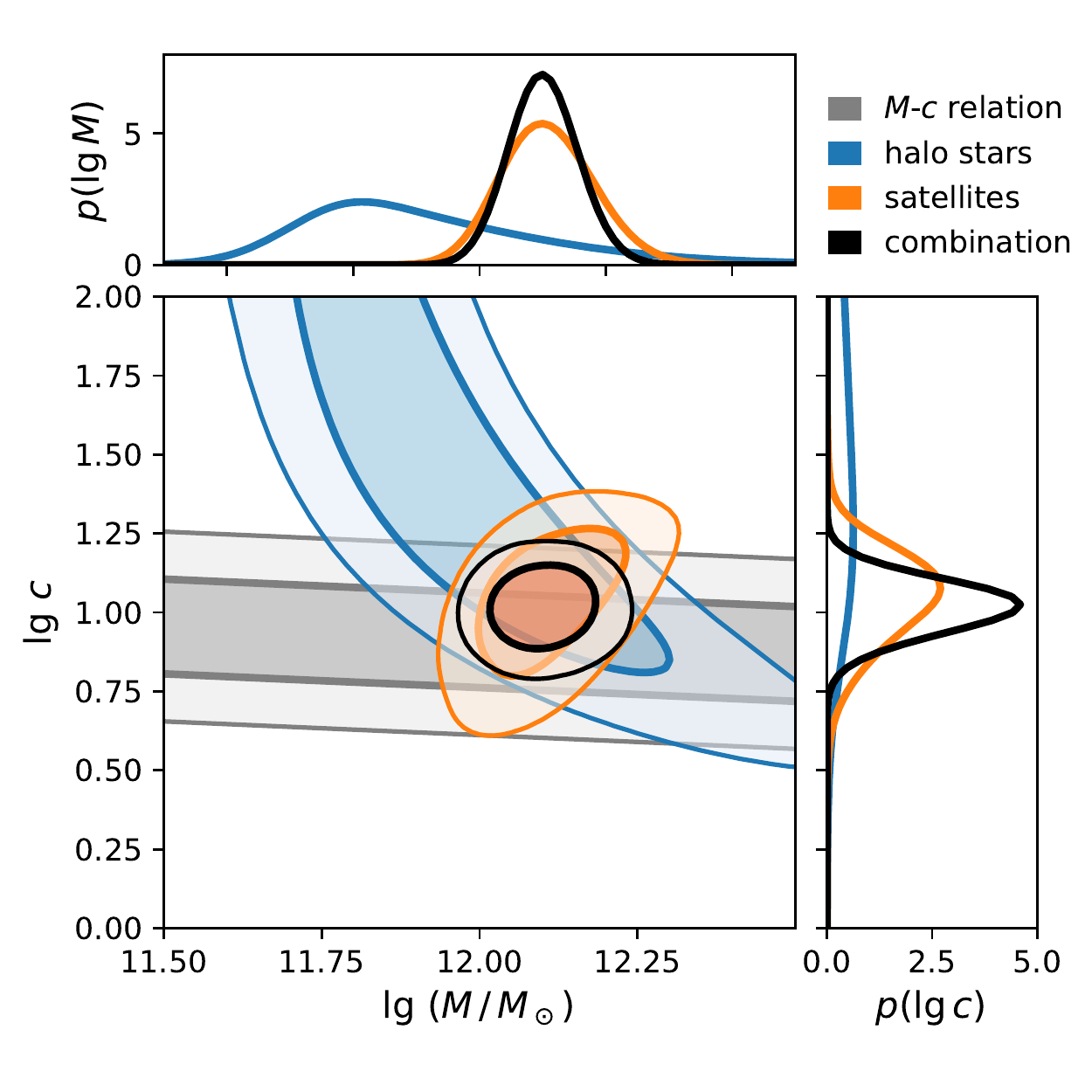}
  \vspace{-4mm}
  \caption{
  Similar to the right panel of \reffig{fig:esti_contour}, but showing additional results
  obtained using the RC from halo stars. The blue (orange) contours show the 1 $\sigma$ and 2 $\sigma$ 
  confidence regions inferred using only the RC from halo stars (satellite kinematics with a flat prior on $\lg c$). 
  The black contours show the results obtained by combining the RC from halo stars and satellite 
  kinematics with the prior based on the $M$-$c$ relation (shown as the gray contours). See text for details.
  }
  \label{fig:esti_combine}
\end{figure}

\reffig{fig:esti_combine} shows the MW halo parameters inferred using
(1) the RC from halo stars, (2) the EAGLE DF for satellite kinematics with a flat prior on $\lg c$, and
(3) a combination of (1) and the EAGLE DF with the prior based on the $M$-$c$ relation.
It can be seen that while the constraints on $\lg M$ and $\lg c$ from (1) are rather loose,
they are approximately orthogonal to those from (2).
In addition, the overlap of these two sets of constraints is in remarkable agreement with the $M$-$c$ 
relation (here taken from the EAGLE simulation but similar to those from other simulations), which
nicely illustrates how the best constraints are obtained using (3). For numerical results,
by combining the RC from halo stars with satellite kinematics, we obtain 
$M=1.27_{-0.15}^{+0.17} \times 10^{12}\msun$ and $c=11.7_{ -2.5}^{ +3.2}$ 
($M=1.26_{-0.15}^{+0.17} \times 10^{12}\msun$ and $c=10.4_{ -1.9}^{ +2.3}$)
for a flat prior on $\lg c$ (the prior based on the $M$-$c$ relation), to be compared with
$M=1.29_{-0.20}^{+0.24} \times 10^{12}\msun$ and $c=11.0_{ -3.3}^{ +4.8}$ 
($M=1.23_{-0.18}^{+0.21} \times 10^{12}\msun$ and $c=9.4_{ -2.1}^{ +2.8}$)
from satellite kinematics alone (see \reftab{tab:esti_result}).
The joint constraints only slightly improve the precision of $M$ 
because satellites are the best tracers of halo mass.
On the other hand, when a flat prior on $\lg c$ is used, the joint constraints
significantly improve the precision of $c$ due to additional constraints 
from halo stars on the inner profile.
Effectively, the joint constraints remove the need for the prior based on the $M$-$c$ relation.

Clearly, the gain from adding a tracer population increases with the precision of
the relevant data and the understanding of the potential systematics.
The use of halo stars as tracers will certainly benefit from {\it Gaia} and its future data release,
as well as other ongoing spectroscopic surveys. These programs can reach further into
the outer halo, and more importantly, they can get rid of the mass-anisotropy degeneracy 
and reduce the substructure contamination (e.g., \citealt{Bird2018})
by directly measuring 3D velocities of halo stars.

\section{Summary and Conclusions} \label{sec:conclusion}

We have estimated the mass and concentration of the MW halo using the kinematic data on 
its satellite galaxies, including the latest measurements from {\it Gaia} DR2. 
Using realistic 6D phase-space DFs of satellite kinematics constructed from cosmological simulations,
we can infer the halo properties efficiently and without bias, and handle the selection function
and measurement errors rigorously in the Bayesian framework.
Applying our DF from the EAGLE simulation to 28 satellites, we obtain an MW halo mass of 
$M=1.23_{-0.18}^{+0.21}\times 10^{12} M_\odot$ and a concentration of $c=9.4_{ -2.1}^{ +2.8}$
with the prior based on the $M$-$c$ relation. The systematic uncertainties in $M$
due to halo-to-halo scatter ($\sim\! 7\%$) and to differences among 
hydrodynamics-based simulations ($\sim\! 5\%$) are small compared to the current statistic error
($\sim\! 17\%$). Due to proper treatment of observational effects, our results are insensitive to
sample selection. In addition, they seem robust against the massive neighbor M31 or the massive
satellite LMC. We recommend the above results as currently the best estimates of the MW mass
and its profile in the outer halo.\footnote{
Note that our estimated concentration is for the total mass profile including baryonic contribution,
and is expected to be slightly higher than the concentration for the dark matter profile.}

Our MW mass estimate is consistent with the latest estimates from various tracers 
(e.g., \citealt{Zhai2018,Sohn2018,Watkins2018,Vasiliev2018,Cautun2019,Fritz2020}, 
see also the review by \citealt{Wang2019}) and, in particular, 
with those using satellite orbital distributions from simulations
\citep{Li2017,Patel2018,Callingham2018}.
However, our estimate is more precise and reliable due to the improved methodology and data. 

Our mass estimate is also in good agreement with
the estimates from the escape velocity of halo stars (e.g., \citealt{Deason2019,Grand2019})
and the timing argument with halo stars \citep{Zaritsky2020}
or nearby galaxies (\citealt{Penarrubia2016,Penarrubia2017}),\footnote{
Note that our estimate should be compared with the total mass of the MW plus the LMC in \citet{Penarrubia2016} and \citet{Penarrubia2017}.
}
which represent completely different approaches to deriving the mass.

In addition, our inferred MW mass profile is consistent with previous measurements
using halo stars \citep{Xue2008,Huang2016, Ablimit2017, Zhai2018}, globular clusters 
\citep{Sohn2018,Watkins2018,Vasiliev2018,Eadie2018}, and satellite galaxies \citep{Cautun2019,Fritz2020}.
Studies using the RC of halo stars usually gave smaller MW mass estimates,
most likely due to biased profile extrapolation to the outer halo.
For example, ignoring the contraction of dark matter profile in the inner halo
would lead to biased profile extrapolation \citep{Cautun2019}.
Because satellites are the proper tracers of the outer halo,
the above issue is irrelevant for our results.
Halo stars are also expected to have larger intrinsic systematics
due to the larger deviation from steady state compared to satellites \citep[e.g.,][]{Wang2017b,Wang2018,Han2019}.

We have also presented results from the SAM-MII DF based on a dark-matter-only simulation.
By comparing both this DF and the EAGLE DF with the observations, we have shown that the 
hydrodynamics-based EAGLE simulation provides a better description of MW satellite kinematics.
Using the EAGLE DF and the associated best-fit MW potential, we have provided much more precise 
estimates of kinematics for those satellites with uncertain measurements, which
may help to better understand their past evolution and the assembly history of the MW.

By comparing the SAM-MII and EAGLE DFs, we find that the former over-represents satellites 
with small radii and velocities, most likely because the gravitational effects of the stellar disk
were not accounted for in the SAM-MII simulation. Such effects include the enhancement of the
tidal field and hence the disruption rate for satellites with small pericenter distances $\rperi$.
The inadequate satellite disruption is likely the main cause of
the earlier reported discrepancy in the velocity anisotropy between the MW satellite system
and the SAM-MII \citep{Cautun2016} or APOSTLE simulation suite \citep{Riley2018}.
We have shown that the differences among hydrodynamics-based simulations may be mimicked by prescribing 
the satellite disruption rate as a function of $\rperi$ in the SAM-MII simulation, which allows us
to estimate the scatter ($\sim\! 5\%$) of halo mass estimates from different hydrodynamics-based simulations.

In the future, the ongoing and planned surveys will increase both the number of tracers in
different populations and the quality of the relevant data, which in turn, will enable us 
to determine the MW halo properties with increasing accuracy. For example,
the number of known satellites may eventually increase by a factor of $\sim 2$--10 \citep{Simon2019}.
The statistical uncertainty decreases as $1/\sqrt{n_\mathrm{sat}}$,
and becomes comparable to the systematic uncertainty when the number of satellites
with complete kinematic data reaches $n_\mathrm{sat} \sim 100$ \citep{Li2019}.
Ultimately, a better understanding of the particular MW formation history and its influence on the mass estimate is required to reduce the systematics.
Note that whereas we have selected the satellites with full kinematic data for convenience of analysis in this study,
our method can treat satellites with incomplete data as well \citep{Li2019}.
In addition, if different tracer populations have independent systematics,
combining multiple tracer populations can further improve the precision by reducing 
the systematic uncertainties. As an illustration, we have combined the RC from halo stars
with satellite kinematics to demonstrate the potential of this approach to improve estimates
of halo properties. Because halo stars and satellites probe different regions of the outer
halo, their combined use effectively removes the need for the prior based on the $M$-$c$ relation.

Compared to satellites, stars and stellar clusters are currently less well understood due to limited 
resolution and various model uncertainties of the simulations. Nevertheless, when we have the proper 
simulations for these tracers, our simulation-based DF method can also apply to e.g., halo stars or 
globular clusters. 
In general, the quality of any DF can be judged based on the Bayesian evidence 
or a direct comparison of the DF with the observed tracer kinematics.
On the other hand, non-parametric methods \citep[e.g.,][]{Bovy2010,Magorrian2014,Han2016b},
which suffer less from model assumptions, might be attractive alternatives for dynamical modeling 
of halo stars or globular clusters when more and better data are available.

\acknowledgments

We thank Marius Cautun, Alis Deason, Carlos S. Frenk, Yang Huang, Lu Li, Chengze Liu, Houjun Mo,  
Zhengyi Shao, Alessandro Sonnenfeld, and Yanqiong Zhang for helpful discussions,
Thomas Callingham and Matthieu Schaller for discussions and for providing the EAGLE halo sample,
and Meng Zhai for providing the data on stellar rotation curves.
We also thank the anonymous referee for constructive criticisms
and helpful suggestions.
This work was supported in part by 
the National Key Basic Research and Development Program of China [2018YFA0404504],
the National Natural Science Foundation of China [11533006, 11621303, 11890691, 11655002, 11873038, 11973032],
the US Department of Energy [DE-FG02-87ER40328 (UM)],
the National Program on Key Basic Research Project [2015CB857003], 
the Science and Technology Commission of Shanghai Municipality [16DZ2260200],
and JSPS Grant-in-Aid for Scientific Research [JP17K14271].
TSL is supported by NASA through Hubble Fellowship grant HF2-51439.001 awarded by the Space Telescope Science Institute, 
which is operated by the Association of Universities for Research in Astronomy, Inc., for NASA, under contract NAS5-26555.

We acknowledge the Virgo Consortium for making their simulation data available. 
The EAGLE simulations were performed using the DiRAC-2 facility at Durham, managed by the ICC, 
and the PRACE facility Curie based in France at TGCC, CEA, Bruy\`{e}res-le-Ch\^{a}tel.

This work made use of the High Performance Computing Resource 
in the Core Facility for Advanced Research Computing at Shanghai Astronomical Observatory,
and the computing facilities at the Department of Astronomy, School of Physics and Astronomy, 
Shanghai Jiao Tong University.

\textit{Software:} 
  Astropy \citep{AstropyCollaboration2013},
  PARSEC \citep{Bressan2012},
  scikit-learn \citep{Pedregosa2011},
  Numpy \citep{Walt2011}, 
  Scipy \citep{Oliphant2007},
  Matplotlib \citep{Hunter2007},
  WebPlotDigitizer,
  adstex

\appendix \label{appendix}
\section{MW satellite properties and posterior kinematics} \label{sec:data_table}

Table \ref{tab:properties} lists the observed properties of those MW satellites used in our study,
including the coordinates, absolute magnitude, distance, line-of-sight velocity, and proper motion.
They are taken from Table A1 (gold sample when possible) of the compilation by 
\citet{Riley2018}. Two additional entries list the posterior proper motion estimates derived from
our EAGLE DF with the prior based on the $M$-$c$ relation (see \refsec{sec:reduced_kin}).

Table \ref{tab:properties2} lists the Galactocentric position and velocity, 
as well as the corresponding uncertainties, 
obtained by Monte Carlo sampling for each satellite (see \refsec{sec:satellites} for detail).
Four additional entries list the posterior kinematics derived in \refsec{sec:reduced_kin}.
These values are listed for reference. We recommend that readers of interest instead use 
the Monte Carlo sample and the corresponding importance weights, 
which are available online at \url{https://github.com/syrte/mw_sats_kin}
and are archived in China-VO (doi:\href{https://doi.org/10.12149/101018}{10.12149/101018}).

\movetabledown=5.3cm
\begin{table*}[p]
\begin{rotatetable}
\begin{minipage}{\textheight}
\caption{Properties of the satellites used in this study: RA ($\alpha$), Dec ($\delta$), absolute magnitude, heliocentric distance, line-of-sight velocity, and proper motion. They are taken from a compilation by \citet{Riley2018}. References for original observations are also given. The last two columns list 
our posterior proper motion estimates. Note that $\mu_{\alpha^{\ast}}\equiv\mu_{\alpha}\cos\delta$.
}
\label{tab:properties}

{\centering
\def\arraystretch{0.95}
\begin{tabular}{lrrrrrrrrrr}
  \midrule\midrule 
    \multicolumn{1}{l}{  Satellite  }&
    \multicolumn{1}{c}{  RA  }&
    \multicolumn{1}{c}{  Dec  }&
    \multicolumn{1}{c}{  $M_V$  }&
    \multicolumn{1}{c}{  $D_{\odot}$  }&
    \multicolumn{1}{c}{  $v_{\odot}$  }&
    \multicolumn{1}{c}{  $\mu_{\alpha^{\ast}}$  }&
    \multicolumn{1}{c}{  $\mu_{\delta}$  }&
    \multicolumn{1}{c}{  Reference  }&
    \multicolumn{1}{c}{  $\mu_{\alpha^{\ast}}^\mathrm{post}$  }&
    \multicolumn{1}{c}{  $\mu_{\delta}^\mathrm{post}$  }\\
    \multicolumn{1}{c}{  }&
    \multicolumn{1}{c}{  [deg]  }&
    \multicolumn{1}{c}{  [deg]  }&
    \multicolumn{1}{c}{  [mag]  }&
    \multicolumn{1}{c}{  [$\kpc$]  }&
    \multicolumn{1}{c}{  [$\kms$]  }&
    \multicolumn{1}{c}{  [$\masyr$]  }&
    \multicolumn{1}{c}{  [$\masyr$]  }&
    \multicolumn{1}{c}{  }&
    \multicolumn{1}{c}{  [$\masyr$]  }&
    \multicolumn{1}{c}{  [$\masyr$]  }\\
  \midrule
 Aquarius II         &  $338.481$  &  $ -9.327$  &  $  -4.36$  &  $ 107.9 \pm   3.3$  &  $ -71.1 \pm   2.5$  &  $-0.491 \pm 0.306$  &  $-0.049 \pm 0.266$  &  [7, 1]           &  $-0.098_{-0.131}^{+0.164}$  &  $-0.223_{-0.191}^{+0.127}$ \\ 
 Bootes I            &  $210.015$  &  $ 14.512$  &  $   -6.3$  &  $    66 \pm     3$  &  $ 102.2 \pm   0.8$  &  $-0.459 \pm 0.041$  &  $-1.064 \pm 0.029$  &  [8, 9, 10, 2]    &  $-0.460_{-0.042}^{+0.043}$  &  $-1.063_{-0.029}^{+0.028}$ \\ 
 Canes Venatici I    &  $202.016$  &  $ 33.559$  &  $   -8.6$  &  $   210 \pm     6$  &  $  30.9 \pm   0.6$  &  $-0.159 \pm   0.1$  &  $-0.067 \pm 0.064$  &  [9, 11, 12, 3]   &  $-0.159_{-0.063}^{+0.062}$  &  $-0.113_{-0.053}^{+0.047}$ \\ 
 Canes Venatici II   &  $194.292$  &  $ 34.321$  &  $   -4.6$  &  $   160 \pm     7$  &  $-128.9 \pm   1.2$  &  $-0.342 \pm 0.238$  &  $-0.473 \pm 0.178$  &  [12, 13, 14, 3]  &  $-0.227_{-0.097}^{+0.125}$  &  $-0.333_{-0.103}^{+0.096}$ \\ 
 Carina I            &  $100.407$  &  $-50.966$  &  $   -8.6$  &  $ 105.6 \pm   5.4$  &  $ 222.9 \pm   0.1$  &  $ 0.495 \pm 0.015$  &  $ 0.143 \pm 0.014$  &  [15, 16, 17, 2]  &  $ 0.493_{-0.014}^{+0.016}$  &  $ 0.143_{-0.014}^{+0.013}$ \\ 
 Coma Berenices I    &  $186.746$  &  $ 23.908$  &  $   -3.8$  &  $    42 \pm   1.5$  &  $  98.1 \pm   0.9$  &  $ 0.546 \pm 0.092$  &  $-1.726 \pm 0.086$  &  [12, 18, 19, 4]  &  $ 0.522_{-0.091}^{+0.093}$  &  $-1.711_{-0.089}^{+0.083}$ \\ 
 Crater II           &  $ 177.31$  &  $-18.413$  &  $   -8.2$  &  $ 117.5 \pm   1.1$  &  $  87.5 \pm   0.4$  &  $-0.246 \pm 0.052$  &  $-0.227 \pm 0.026$  &  [20, 21, 1]      &  $-0.247_{-0.052}^{+0.050}$  &  $-0.226_{-0.025}^{+0.025}$ \\ 
 Draco I             &  $ 260.06$  &  $ 57.965$  &  $  -8.75$  &  $    76 \pm     6$  &  $  -291 \pm   0.1$  &  $-0.019 \pm 0.009$  &  $-0.145 \pm  0.01$  &  [22, 23, 2]      &  $-0.019_{-0.009}^{+0.009}$  &  $-0.145_{-0.010}^{+0.010}$ \\ 
 Fornax              &  $ 39.962$  &  $-34.511$  &  $  -13.4$  &  $   147 \pm     9$  &  $  55.3 \pm   0.1$  &  $ 0.376 \pm 0.003$  &  $-0.413 \pm 0.003$  &  [17, 24, 2]      &  $ 0.376_{-0.003}^{+0.003}$  &  $-0.413_{-0.003}^{+0.003}$ \\ 
 Grus I              &  $344.176$  &  $-50.163$  &  $   -3.4$  &  $ 120.2 \pm  11.1$  &  $-140.5 \pm     2$  &  $ -0.25 \pm  0.16$  &  $ -0.47 \pm  0.23$  &  [25, 26, 5]      &  $-0.045_{-0.101}^{+0.106}$  &  $-0.453_{-0.129}^{+0.142}$ \\ 
 Hercules            &  $247.763$  &  $ 12.787$  &  $   -6.6$  &  $   132 \pm     6$  &  $  45.2 \pm  1.09$  &  $-0.297 \pm 0.123$  &  $-0.329 \pm   0.1$  &  [27, 28, 3]      &  $-0.284_{-0.098}^{+0.103}$  &  $-0.314_{-0.084}^{+0.087}$ \\ 
 Horologium I        &  $ 43.882$  &  $-54.119$  &  $   -3.5$  &  $    79 \pm     7$  &  $ 112.8 \pm  2.55$  &  $  0.95 \pm  0.07$  &  $ -0.55 \pm  0.06$  &  [29, 5]          &  $ 0.935_{-0.070}^{+0.069}$  &  $-0.542_{-0.058}^{+0.057}$ \\ 
 Hydra II            &  $185.425$  &  $-31.985$  &  $   -4.8$  &  $   151 \pm     8$  &  $ 303.1 \pm   1.4$  &  $-0.417 \pm 0.402$  &  $ 0.179 \pm 0.339$  &  [30, 31, 32, 1]  &  $-0.228_{-0.115}^{+0.162}$  &  $-0.094_{-0.112}^{+0.114}$ \\ 
 Leo I               &  $152.122$  &  $ 12.313$  &  $ -12.03$  &  $ 258.2 \pm   9.5$  &  $ 282.5 \pm   0.1$  &  $-0.097 \pm 0.056$  &  $-0.091 \pm 0.047$  &  [33, 34, 2]      &  $-0.071_{-0.039}^{+0.041}$  &  $-0.123_{-0.037}^{+0.035}$ \\ 
 Leo II              &  $ 168.37$  &  $ 22.152$  &  $   -9.6$  &  $   233 \pm    15$  &  $  78.5 \pm   0.6$  &  $-0.064 \pm 0.057$  &  $ -0.21 \pm 0.054$  &  [35, 36, 2]      &  $-0.071_{-0.046}^{+0.045}$  &  $-0.207_{-0.045}^{+0.046}$ \\ 
 Leo IV              &  $173.233$  &  $  -0.54$  &  $  -4.97$  &  $   154 \pm     5$  &  $ 132.3 \pm   1.4$  &  $ -0.59 \pm 0.534$  &  $-0.449 \pm 0.362$  &  [12, 37, 3]      &  $-0.157_{-0.145}^{+0.125}$  &  $-0.270_{-0.147}^{+0.135}$ \\ 
 Leo V               &  $172.784$  &  $  2.222$  &  $   -4.4$  &  $   173 \pm     5$  &  $ 172.1 \pm   2.2$  &  $-0.097 \pm  0.56$  &  $-0.628 \pm 0.307$  &  [38, 39, 3]      &  $-0.128_{-0.093}^{+0.096}$  &  $-0.272_{-0.101}^{+0.125}$ \\ 
 LMC                 &  $ 80.894$  &  $-69.756$  &  $  -18.1$  &  $    51 \pm     2$  &  $ 262.2 \pm   3.4$  &  $  1.85 \pm  0.03$  &  $  0.24 \pm  0.03$  &  [40, 2]          &  $ 1.845_{-0.031}^{+0.031}$  &  $ 0.241_{-0.030}^{+0.027}$ \\ 
 Pisces II           &  $344.634$  &  $  5.955$  &  $   -4.1$  &  $   183 \pm    15$  &  $-226.5 \pm   2.7$  &  $-0.108 \pm 0.647$  &  $-0.586 \pm 0.502$  &  [14, 30, 3]      &  $ 0.096_{-0.101}^{+0.113}$  &  $-0.215_{-0.093}^{+0.111}$ \\ 
 Sculptor            &  $ 15.039$  &  $-33.709$  &  $  -10.7$  &  $  83.9 \pm   1.5$  &  $ 111.4 \pm   0.1$  &  $ 0.082 \pm 0.005$  &  $-0.131 \pm 0.004$  &  [17, 41, 2]      &  $ 0.082_{-0.005}^{+0.005}$  &  $-0.131_{-0.004}^{+0.004}$ \\ 
 Segue 2             &  $ 34.817$  &  $ 20.175$  &  $   -2.5$  &  $  36.6 \pm  2.45$  &  $ -40.2 \pm   0.9$  &  $  1.01 \pm  0.14$  &  $ -0.48 \pm  0.18$  &  [42, 43, 6]      &  $ 1.002_{-0.149}^{+0.152}$  &  $-0.447_{-0.203}^{+0.194}$ \\ 
 Sextans             &  $153.268$  &  $  -1.62$  &  $   -9.3$  &  $  92.5 \pm   2.2$  &  $ 224.2 \pm   0.1$  &  $-0.496 \pm 0.025$  &  $ 0.077 \pm  0.02$  &  [17, 44, 2]      &  $-0.490_{-0.024}^{+0.025}$  &  $ 0.072_{-0.021}^{+0.019}$ \\ 
 SMC                 &  $ 13.187$  &  $-72.829$  &  $  -16.8$  &  $    64 \pm     4$  &  $ 145.6 \pm   0.6$  &  $ 0.797 \pm  0.03$  &  $ -1.22 \pm  0.03$  &  [40, 2]          &  $ 0.796_{-0.030}^{+0.028}$  &  $-1.214_{-0.031}^{+0.030}$ \\ 
 Tucana II           &  $ 343.06$  &  $ -58.57$  &  $   -3.9$  &  $  57.5 \pm   5.3$  &  $-129.1 \pm   3.5$  &  $  0.91 \pm  0.06$  &  $ -1.16 \pm  0.08$  &  [26, 45, 5]      &  $ 0.905_{-0.059}^{+0.058}$  &  $-1.152_{-0.079}^{+0.075}$ \\ 
 Ursa Major I        &  $158.685$  &  $ 51.926$  &  $  -6.75$  &  $  97.3 \pm  5.85$  &  $ -55.3 \pm   1.4$  &  $-0.659 \pm 0.093$  &  $-0.635 \pm 0.131$  &  [12, 46, 4]      &  $-0.590_{-0.080}^{+0.084}$  &  $-0.616_{-0.112}^{+0.114}$ \\ 
 Ursa Major II       &  $132.874$  &  $ 63.133$  &  $   -3.9$  &  $  34.7 \pm   2.1$  &  $-116.5 \pm   1.9$  &  $ 1.661 \pm 0.053$  &  $ -1.87 \pm 0.065$  &  [12, 47, 4]      &  $ 1.656_{-0.052}^{+0.052}$  &  $-1.868_{-0.066}^{+0.065}$ \\ 
 Ursa Minor          &  $227.242$  &  $ 67.222$  &  $   -8.4$  &  $    76 \pm     4$  &  $-246.9 \pm   0.1$  &  $-0.182 \pm  0.01$  &  $ 0.074 \pm 0.008$  &  [48, 2]          &  $-0.181_{-0.010}^{+0.010}$  &  $ 0.074_{-0.008}^{+0.008}$ \\ 
 Willman 1           &  $162.341$  &  $ 51.053$  &  $   -2.7$  &  $    38 \pm     7$  &  $ -12.8 \pm     1$  &  $ 0.382 \pm 0.119$  &  $-1.152 \pm 0.216$  &  [9, 49, 4]       &  $ 0.372_{-0.117}^{+0.117}$  &  $-1.167_{-0.203}^{+0.210}$ \\ 
  \midrule\midrule
\end{tabular}\\
}
\vspace{0.5ex}
\textbf{References}. [1] \citet{Kallivayalil2018}; [2] \citet{GaiaCollaboration2018}; [3] \citet{Fritz2018a}; [4] \citet{Simon2018}; [5] \citet{Pace2018b}; [6] \citet{Massari2018}; [7] \citet{2016MNRAS.463..712T}; [8] \citet{2006ApJ...653L.109D}; [9] \citet{2008ApJ...684.1075M}; [10] \citet{2011ApJ...736..146K}; [11] \citet{2008ApJ...674L..81K}; [12] \citet{2007ApJ...670..313S}; [13] \citet{2008ApJ...675L..73G}; [14] \citet{2012ApJ...756...79S}; [15] \citet{2015AJ....150...90K}; [16] \citet{2014MNRAS.444.3139M}; [17] \citet{2009AJ....137.3100W}; [18] \citet{2010AJ....140..138M}; [19] \citet{2009ApJ...695L..83M}; [20] \citet{2017ApJ...839...20C}; [21] \citet{2016MNRAS.459.2370T}; [22] \citet{2015MNRAS.448.2717W}; [23] \citet{2004AJ....127..861B}; [24] \citet{2009AJ....138..459P}; [25] \citet{2015ApJ...805..130K}; [26] \citet{2016ApJ...819...53W}; [27] \citet{2009A&A...506.1147A}; [28] \citet{2012ApJ...756..121M}; [29] \citet{2015ApJ...811...62K}; [30] \citet{2015ApJ...810...56K}; [31] \citet{2016AJ....151..118V}; [32] \citet{2015ApJ...804L...5M}; [33] \citet{2008ApJ...675..201M}; [34] \citet{2014PASP..126..616S}; [35] \citet{2017ApJ...836..202S}; [36] \citet{2005MNRAS.360..185B}; [37] \citet{2009ApJ...699L.125M}; [38] \citet{2017MNRAS.467..573C}; [39] \citet{2017ApJ...845L..10M}; [40] \citet{2012AJ....144....4M}; [41] \citet{2015MNRAS.454.1509M}; [42] \citet{2013AJ....146...94B}; [43] \citet{2013ApJ...770...16K}; [44] \citet{2017MNRAS.467..208O}; [45] \citet{2015ApJ...807...50B}; [46] \citet{2013ApJ...767...62G}; [47] \citet{2012ApJ...752...42D}; [48] \citet{2002AJ....124.3222B}; [49] \citet{2011AJ....142..128W}.

\end{minipage}
\end{rotatetable}
\end{table*}

\movetabledown=5.3cm
\begin{table*}[p]
\begin{rotatetable}
\begin{minipage}{\textheight}
\caption{Galactocentric positions and velocities of the satellites. The last four columns
list our estimated posterior kinematics. All values are in terms of the median and the $1\sigma$ uncertainty.
}

\centering
\def\arraystretch{1.2}
\begin{tabular}{lrrrrrrrrrr}
  \midrule \midrule
    \multicolumn{1}{l}{  Satellite  }&
    \multicolumn{1}{c}{  $r$  }&
    \multicolumn{1}{c}{  $\theta$  }&
    \multicolumn{1}{c}{  $\phi$  }&
    \multicolumn{1}{c}{  $v_r$  }&
    \multicolumn{1}{c}{  $v_{\theta}$  }&
    \multicolumn{1}{c}{  $v_{\phi}$  }&
    \multicolumn{1}{c}{  $r^\mathrm{post}$  }&
    \multicolumn{1}{c}{  $v^\mathrm{post}_r$  }&
    \multicolumn{1}{c}{  $v^\mathrm{post}_{\theta}$  }&
    \multicolumn{1}{c}{  $v^\mathrm{post}_{\phi}$  }\\
    \multicolumn{1}{c}{  }&
    \multicolumn{1}{c}{  [kpc]  }&
    \multicolumn{1}{c}{  [deg]  }&
    \multicolumn{1}{c}{  [deg]  }&
    \multicolumn{1}{c}{  [$\kms$]  }&
    \multicolumn{1}{c}{  [$\kms$]  }&
    \multicolumn{1}{c}{  [$\kms$]  }&
    \multicolumn{1}{c}{  [$\kpc$]  }&
    \multicolumn{1}{c}{  [$\kms$]  }&
    \multicolumn{1}{c}{  [$\kms$]  }&
    \multicolumn{1}{c}{  [$\kms$]  }\\
  \midrule
 Aquarius II         &  $ 105.3_{  -3.5}^{  +3.2}$  &  $ 145.0_{  -0.1}^{  +0.1}$  &  $  61.7_{  -0.2}^{  +0.2}$  &  $  34.2_{ -11.4}^{ +11.7}$  &  $-372.1_{-145.4}^{+152.8}$  &  $ -62.5_{-138.8}^{+146.8}$  &  $ 105.0_{  -3.3}^{  +3.4}$  &  $  44.4_{  -6.7}^{  +6.9}$  &  $-149.0_{ -55.4}^{ +94.3}$  &  $  -6.4_{ -91.2}^{ +88.3}$ \\ 
 Bootes I            &  $  63.7_{  -2.9}^{  +2.9}$  &  $  13.6_{  -0.3}^{  +0.3}$  &  $ 357.1_{  -0.1}^{  +0.1}$  &  $  95.8_{  -1.6}^{  +1.7}$  &  $ 123.1_{ -12.2}^{ +11.9}$  &  $ -94.8_{ -19.7}^{ +19.7}$  &  $  63.5_{  -2.9}^{  +2.9}$  &  $  95.8_{  -1.6}^{  +1.6}$  &  $ 122.7_{ -12.0}^{ +11.9}$  &  $ -93.5_{ -19.5}^{ +19.1}$ \\ 
 Canes Venatici I    &  $ 210.0_{  -6.1}^{  +5.9}$  &  $   9.8_{  -0.0}^{  +0.0}$  &  $  86.0_{  -0.4}^{  +0.4}$  &  $  83.3_{  -3.4}^{  +3.6}$  &  $  89.2_{ -81.2}^{ +82.7}$  &  $  76.2_{ -83.2}^{ +88.9}$  &  $ 209.6_{  -6.5}^{  +5.7}$  &  $  82.2_{  -2.3}^{  +2.3}$  &  $  53.1_{ -54.9}^{ +57.7}$  &  $  46.6_{ -56.5}^{ +57.5}$ \\ 
 Canes Venatici II   &  $ 160.6_{  -6.8}^{  +7.0}$  &  $   8.8_{  -0.1}^{  +0.1}$  &  $ 130.4_{  -0.6}^{  +0.6}$  &  $ -93.3_{  -8.9}^{  +9.2}$  &  $-144.0_{-142.8}^{+136.8}$  &  $ 134.3_{-182.3}^{+187.0}$  &  $ 158.6_{  -6.3}^{  +7.5}$  &  $ -94.2_{  -4.1}^{  +4.3}$  &  $ -47.3_{ -71.5}^{ +74.2}$  &  $  39.1_{ -93.3}^{ +69.4}$ \\ 
 Carina I            &  $ 107.5_{  -5.3}^{  +5.1}$  &  $ 111.8_{  -0.0}^{  +0.0}$  &  $ 255.4_{  -0.2}^{  +0.2}$  &  $  -4.2_{  -0.6}^{  +0.5}$  &  $-173.0_{ -14.1}^{ +14.8}$  &  $ -27.4_{  -7.0}^{  +6.8}$  &  $ 106.1_{  -5.1}^{  +5.1}$  &  $  -4.1_{  -0.6}^{  +0.5}$  &  $-169.2_{ -14.1}^{ +15.2}$  &  $ -27.4_{  -7.0}^{  +6.6}$ \\ 
 Coma Berenices I    &  $  43.2_{  -1.5}^{  +1.5}$  &  $  14.8_{  -0.3}^{  +0.4}$  &  $ 201.8_{  -0.6}^{  +0.6}$  &  $  26.5_{  -3.6}^{  +3.5}$  &  $-265.9_{ -19.6}^{ +18.3}$  &  $ 103.3_{ -21.3}^{ +22.5}$  &  $  43.0_{  -1.4}^{  +1.5}$  &  $  27.3_{  -3.5}^{  +3.4}$  &  $-261.3_{ -19.4}^{ +17.5}$  &  $  99.6_{ -21.0}^{ +22.7}$ \\ 
 Crater II           &  $ 116.4_{  -1.1}^{  +1.1}$  &  $  47.5_{  -0.0}^{  +0.0}$  &  $ 277.7_{  -0.1}^{  +0.1}$  &  $ -83.7_{  -1.9}^{  +1.9}$  &  $  -1.7_{ -16.3}^{ +17.8}$  &  $ -30.6_{ -27.0}^{ +26.5}$  &  $ 116.4_{  -1.1}^{  +1.1}$  &  $ -83.7_{  -1.9}^{  +1.9}$  &  $  -1.7_{ -16.2}^{ +17.8}$  &  $ -30.5_{ -27.2}^{ +26.4}$ \\ 
 Draco I             &  $  76.2_{  -6.1}^{  +6.0}$  &  $  55.3_{  -0.0}^{  +0.0}$  &  $  93.8_{  -0.6}^{  +0.7}$  &  $ -88.0_{  -0.4}^{  +0.4}$  &  $ 122.4_{  -3.4}^{  +3.2}$  &  $ -45.7_{  -6.1}^{  +5.2}$  &  $  75.8_{  -6.2}^{  +6.0}$  &  $ -88.0_{  -0.4}^{  +0.4}$  &  $ 122.4_{  -3.4}^{  +3.2}$  &  $ -45.4_{  -6.1}^{  +5.2}$ \\ 
 Fornax              &  $ 149.0_{  -9.1}^{  +8.9}$  &  $ 153.9_{  -0.1}^{  +0.1}$  &  $ 230.9_{  -0.4}^{  +0.3}$  &  $ -41.2_{  -0.2}^{  +0.2}$  &  $-106.3_{ -18.0}^{ +18.1}$  &  $ 120.3_{ -15.9}^{ +15.9}$  &  $ 144.4_{  -8.1}^{  +8.4}$  &  $ -41.2_{  -0.1}^{  +0.2}$  &  $ -97.0_{ -16.9}^{ +15.9}$  &  $ 112.0_{ -14.6}^{ +15.1}$ \\ 
 Grus I              &  $ 115.9_{ -10.7}^{ +11.4}$  &  $ 151.6_{  -0.3}^{  +0.3}$  &  $ 335.5_{  -0.4}^{  +0.3}$  &  $-202.7_{  -6.2}^{  +5.9}$  &  $-183.7_{-115.3}^{+102.6}$  &  $ 109.5_{-122.3}^{+117.1}$  &  $ 110.1_{ -11.0}^{ +11.3}$  &  $-195.6_{  -4.2}^{  +3.7}$  &  $ -84.4_{ -53.3}^{ +63.4}$  &  $  51.0_{ -75.5}^{ +70.8}$ \\ 
 Hercules            &  $ 126.3_{  -5.8}^{  +6.0}$  &  $  51.3_{  -0.1}^{  +0.1}$  &  $  30.9_{  -0.1}^{  +0.1}$  &  $ 150.6_{  -3.3}^{  +3.0}$  &  $ -11.4_{ -71.2}^{ +74.1}$  &  $ -56.9_{ -63.5}^{ +66.1}$  &  $ 125.5_{  -5.8}^{  +5.8}$  &  $ 150.8_{  -2.8}^{  +2.6}$  &  $  -3.8_{ -61.2}^{ +60.5}$  &  $ -43.3_{ -52.3}^{ +57.0}$ \\ 
 Horologium I        &  $  79.4_{  -6.5}^{  +6.6}$  &  $ 144.4_{  -0.1}^{  +0.1}$  &  $ 260.9_{  -0.9}^{  +0.8}$  &  $ -32.5_{  -3.5}^{  +3.5}$  &  $-212.5_{ -42.5}^{ +41.2}$  &  $  -8.0_{ -23.8}^{ +22.9}$  &  $  75.8_{  -6.0}^{  +6.2}$  &  $ -32.6_{  -3.5}^{  +3.6}$  &  $-187.3_{ -36.7}^{ +32.5}$  &  $  -6.9_{ -22.8}^{ +21.2}$ \\ 
 Hydra II            &  $ 148.2_{  -8.0}^{  +8.2}$  &  $  58.9_{  -0.0}^{  +0.0}$  &  $ 292.4_{  -0.2}^{  +0.2}$  &  $ 130.6_{ -15.5}^{ +14.3}$  &  $-194.2_{-243.7}^{+235.0}$  &  $-214.0_{-278.7}^{+299.8}$  &  $ 146.1_{  -6.6}^{  +8.4}$  &  $ 119.5_{  -6.2}^{  +4.1}$  &  $ -30.3_{ -81.0}^{ +81.0}$  &  $ -21.7_{ -81.9}^{+103.6}$ \\ 
 Leo I               &  $ 262.1_{  -9.1}^{  +9.3}$  &  $  41.7_{  -0.0}^{  +0.0}$  &  $ 224.2_{  -0.1}^{  +0.1}$  &  $ 169.5_{  -1.8}^{  +1.9}$  &  $  12.5_{ -66.1}^{ +69.4}$  &  $-118.8_{ -61.9}^{ +59.4}$  &  $ 260.9_{  -9.1}^{  +9.0}$  &  $ 168.1_{  -1.4}^{  +1.3}$  &  $   0.0_{ -47.3}^{ +47.7}$  &  $ -66.6_{ -44.4}^{ +42.9}$ \\ 
 Leo II              &  $ 235.5_{ -14.7}^{ +14.8}$  &  $  24.1_{  -0.1}^{  +0.1}$  &  $ 217.3_{  -0.2}^{  +0.2}$  &  $  19.1_{  -2.2}^{  +2.2}$  &  $ -24.7_{ -64.7}^{ +66.3}$  &  $  15.8_{ -62.7}^{ +61.5}$  &  $ 232.4_{ -13.9}^{ +14.6}$  &  $  19.5_{  -1.7}^{  +1.8}$  &  $ -16.8_{ -49.4}^{ +50.6}$  &  $   8.2_{ -50.5}^{ +49.7}$ \\ 
 Leo IV              &  $ 154.4_{  -4.8}^{  +5.1}$  &  $  33.8_{  -0.0}^{  +0.0}$  &  $ 260.2_{  -0.2}^{  +0.2}$  &  $  12.6_{ -19.1}^{ +19.8}$  &  $ 291.5_{-302.7}^{+342.6}$  &  $-171.8_{-349.3}^{+345.5}$  &  $ 153.5_{  -4.3}^{  +5.0}$  &  $   1.5_{  -5.8}^{  +5.6}$  &  $  25.6_{-112.4}^{ +86.9}$  &  $  -2.4_{-104.1}^{+103.3}$ \\ 
 Leo V               &  $ 173.7_{  -4.9}^{  +5.1}$  &  $  31.9_{  -0.0}^{  +0.0}$  &  $ 257.1_{  -0.1}^{  +0.1}$  &  $  39.7_{ -20.1}^{ +19.8}$  &  $ 231.6_{-354.1}^{+344.8}$  &  $ 250.6_{-383.7}^{+390.6}$  &  $ 173.1_{  -4.7}^{  +4.2}$  &  $  48.1_{  -3.5}^{  +4.8}$  &  $  28.4_{-100.5}^{ +70.4}$  &  $  18.5_{ -82.8}^{ +82.9}$ \\ 
 LMC                 &  $  50.3_{  -1.9}^{  +2.0}$  &  $ 123.3_{  -0.0}^{  +0.0}$  &  $ 269.3_{  -0.4}^{  +0.4}$  &  $  63.0_{  -3.6}^{  +3.4}$  &  $-310.3_{ -18.5}^{ +18.7}$  &  $ -41.4_{  -8.7}^{  +8.6}$  &  $  49.2_{  -1.9}^{  +1.8}$  &  $  62.7_{  -3.5}^{  +3.6}$  &  $-298.4_{ -16.9}^{ +17.0}$  &  $ -38.0_{  -8.3}^{  +8.2}$ \\ 
 Pisces II           &  $ 182.1_{ -15.1}^{ +15.3}$  &  $ 137.4_{  -0.0}^{  +0.0}$  &  $  83.1_{  -0.3}^{  +0.3}$  &  $ -79.6_{ -23.7}^{ +24.7}$  &  $ 139.0_{-461.8}^{+507.3}$  &  $-347.3_{-497.7}^{+533.7}$  &  $ 174.2_{ -12.8}^{ +14.7}$  &  $ -66.1_{  -4.2}^{  +6.6}$  &  $  13.3_{ -85.0}^{ +77.5}$  &  $   5.3_{ -95.4}^{ +82.8}$ \\ 
 Sculptor            &  $  84.0_{  -1.4}^{  +1.4}$  &  $ 172.5_{  -0.0}^{  +0.0}$  &  $ 240.3_{  -0.6}^{  +0.6}$  &  $  74.9_{  -0.2}^{  +0.2}$  &  $ 170.3_{  -2.0}^{  +1.9}$  &  $ -72.7_{  -3.1}^{  +2.7}$  &  $  84.0_{  -1.4}^{  +1.4}$  &  $  74.9_{  -0.2}^{  +0.2}$  &  $ 170.3_{  -2.0}^{  +1.9}$  &  $ -72.8_{  -3.1}^{  +2.7}$ \\ 
 Segue 2             &  $  42.7_{  -2.5}^{  +2.4}$  &  $ 121.9_{  -0.4}^{  +0.3}$  &  $ 156.1_{  -0.3}^{  +0.4}$  &  $  52.7_{  -4.0}^{  +4.1}$  &  $ -70.2_{ -29.3}^{ +31.4}$  &  $ -36.4_{ -29.8}^{ +30.1}$  &  $  42.4_{  -2.6}^{  +2.5}$  &  $  52.8_{  -4.0}^{  +4.2}$  &  $ -73.9_{ -30.7}^{ +33.1}$  &  $ -40.2_{ -29.8}^{ +31.8}$ \\ 
 Sextans             &  $  95.6_{  -2.2}^{  +2.3}$  &  $  49.3_{  -0.0}^{  +0.0}$  &  $ 237.8_{  -0.1}^{  +0.1}$  &  $  81.5_{  -0.9}^{  +0.8}$  &  $   2.9_{ -11.5}^{ +10.9}$  &  $-262.5_{ -10.3}^{ +10.2}$  &  $  94.9_{  -2.2}^{  +2.4}$  &  $  81.2_{  -0.8}^{  +0.8}$  &  $   0.8_{ -10.8}^{ +10.5}$  &  $-257.9_{  -9.7}^{  +9.5}$ \\ 
 SMC                 &  $  61.1_{  -3.7}^{  +4.0}$  &  $ 136.9_{  -0.1}^{  +0.1}$  &  $ 293.1_{  -0.6}^{  +0.6}$  &  $  -5.7_{  -1.4}^{  +1.4}$  &  $-243.8_{ -25.6}^{ +25.9}$  &  $ -66.7_{ -16.4}^{ +15.8}$  &  $  59.2_{  -3.8}^{  +3.6}$  &  $  -6.0_{  -1.4}^{  +1.3}$  &  $-230.3_{ -24.5}^{ +23.6}$  &  $ -59.8_{ -15.2}^{ +14.6}$ \\ 
 Tucana II           &  $  53.7_{  -5.5}^{  +5.4}$  &  $ 148.1_{  -0.5}^{  +0.6}$  &  $ 319.1_{  -1.2}^{  +0.9}$  &  $-187.4_{  -3.9}^{  +4.2}$  &  $  49.1_{ -20.7}^{ +20.8}$  &  $-208.2_{ -40.6}^{ +38.4}$  &  $  51.2_{  -4.8}^{  +4.6}$  &  $-187.7_{  -3.9}^{  +4.3}$  &  $  53.7_{ -20.0}^{ +18.9}$  &  $-189.4_{ -32.4}^{ +30.8}$ \\ 
 Ursa Major I        &  $ 101.9_{  -6.0}^{  +5.9}$  &  $  39.0_{  -0.2}^{  +0.2}$  &  $ 161.9_{  -0.1}^{  +0.1}$  &  $  10.9_{  -3.4}^{  +3.4}$  &  $ 175.0_{ -48.8}^{ +50.9}$  &  $ 163.3_{ -60.2}^{ +62.9}$  &  $  98.5_{  -5.3}^{  +5.8}$  &  $   8.8_{  -3.0}^{  +2.9}$  &  $ 148.1_{ -41.6}^{ +41.6}$  &  $ 127.5_{ -46.4}^{ +46.8}$ \\ 
 Ursa Major II       &  $  41.0_{  -2.1}^{  +2.1}$  &  $  58.8_{  -0.3}^{  +0.4}$  &  $ 158.6_{  -0.3}^{  +0.3}$  &  $ -57.6_{  -2.2}^{  +2.5}$  &  $-276.0_{ -21.2}^{ +22.1}$  &  $  29.9_{ -18.7}^{ +16.7}$  &  $  40.4_{  -2.2}^{  +2.0}$  &  $ -57.7_{  -2.3}^{  +2.4}$  &  $-269.2_{ -21.0}^{ +22.6}$  &  $  25.3_{ -18.2}^{ +16.8}$ \\ 
 Ursa Minor          &  $  77.7_{  -4.1}^{  +3.9}$  &  $  46.5_{  -0.1}^{  +0.1}$  &  $ 112.9_{  -0.4}^{  +0.4}$  &  $ -71.7_{  -0.4}^{  +0.4}$  &  $ 135.4_{  -3.4}^{  +3.3}$  &  $ -14.3_{  -4.9}^{  +5.1}$  &  $  77.5_{  -4.1}^{  +3.9}$  &  $ -71.8_{  -0.4}^{  +0.4}$  &  $ 135.4_{  -3.3}^{  +3.3}$  &  $ -14.5_{  -4.9}^{  +5.1}$ \\ 
 Willman 1           &  $  42.6_{  -6.9}^{  +7.0}$  &  $  41.9_{  -1.3}^{  +1.8}$  &  $ 164.6_{  -0.7}^{  +0.9}$  &  $  13.4_{  -4.2}^{  +3.8}$  &  $-120.2_{ -41.1}^{ +37.0}$  &  $ -96.6_{ -43.4}^{ +47.5}$  &  $  40.3_{  -6.6}^{  +6.8}$  &  $  13.3_{  -4.1}^{  +3.8}$  &  $-109.8_{ -37.5}^{ +35.3}$  &  $-103.9_{ -39.7}^{ +42.9}$ \\ 
  \midrule \midrule
\end{tabular}
\label{tab:properties2}
\end{minipage}
\end{rotatetable}
\end{table*}

\section{Inferred MW mass profile} \label{sec:profile_table}
\reftab{tab:esti_profile} presents the MW mass profile inferred from the EAGLE DF.
The corresponding halo parameters are given in \reftab{tab:esti_result}.
Our satellite sample covers $40<r<280\,\kpc$. Entries outside this range are for reference only.
Similar to the stellar rotation curves, 
these mass profiles can also be used to constrain MW mass models with multiple components
(e.g., bulge, stellar disk, gas disk, and dark matter).
Measurements at different radii should not be taken as independent. Instead,
the covariance between different radii should be taken into account as done in \refeqn{eqn:RC}.
The covariance matrix is provided online.

\begin{table}[htbp]
\caption{
  MW mass profile inferred from the EAGLE DF.
  Our \textit{recommendation} is highlighted.
}
\centering
\def\arraystretch{1.2}
\begin{tabular}{cc>{\columncolor[gray]{.85}}ccc}
\midrule \midrule 
    & \multicolumn{2}{c}{Satellites} & \multicolumn{2}{c}{Satellites + Halo Stars}\\
\cmidrule(lr){2-3} \cmidrule(lr){4-5}
    & {flat prior} & \multicolumn{1}{c}{$M$-$c$ relat.} & {flat prior} & \multicolumn{1}{c}{$M$-$c$ relat. }\\
\midrule
$r$ & $M(<r)$ & \multicolumn{1}{c}{$M(<r)$} & $M(<r)$ & $M(<r)$\\
$[\kpc]$ & $[10^{12}\msun]$ & \multicolumn{1}{c}{$[10^{12}\msun]$} & $[10^{12}\msun]$ & $[10^{12}\msun]$\\
\midrule
    30 & $0.25_{-0.06}^{+0.08}$ & $0.22_{-0.04}^{+0.05}$ & $0.25_{-0.04}^{+0.05}$ & $0.24_{-0.03}^{+0.04}$ \\ 
    40 & $0.34_{-0.08}^{+0.10}$ & $0.30_{-0.05}^{+0.06}$ & $0.35_{-0.05}^{+0.05}$ & $0.32_{-0.04}^{+0.05}$ \\ 
    50 & $0.42_{-0.09}^{+0.11}$ & $0.39_{-0.06}^{+0.07}$ & $0.43_{-0.06}^{+0.06}$ & $0.41_{-0.05}^{+0.05}$ \\ 
    60 & $0.50_{-0.10}^{+0.12}$ & $0.46_{-0.07}^{+0.08}$ & $0.51_{-0.06}^{+0.07}$ & $0.49_{-0.05}^{+0.06}$ \\ 
    80 & $0.64_{-0.12}^{+0.13}$ & $0.60_{-0.08}^{+0.10}$ & $0.65_{-0.07}^{+0.08}$ & $0.62_{-0.07}^{+0.07}$ \\ 
   100 & $0.76_{-0.12}^{+0.15}$ & $0.72_{-0.10}^{+0.11}$ & $0.77_{-0.08}^{+0.08}$ & $0.74_{-0.08}^{+0.08}$ \\ 
   125 & $0.89_{-0.13}^{+0.16}$ & $0.85_{-0.11}^{+0.12}$ & $0.89_{-0.09}^{+0.10}$ & $0.87_{-0.09}^{+0.10}$ \\ 
   150 & $1.00_{-0.14}^{+0.17}$ & $0.96_{-0.12}^{+0.13}$ & $1.00_{-0.10}^{+0.11}$ & $0.99_{-0.10}^{+0.10}$ \\ 
   175 & $1.10_{-0.15}^{+0.18}$ & $1.06_{-0.13}^{+0.14}$ & $1.10_{-0.11}^{+0.12}$ & $1.09_{-0.11}^{+0.12}$ \\ 
   200 & $1.19_{-0.16}^{+0.19}$ & $1.15_{-0.13}^{+0.15}$ & $1.19_{-0.12}^{+0.13}$ & $1.18_{-0.11}^{+0.13}$ \\ 
   225 & $1.27_{-0.16}^{+0.20}$ & $1.23_{-0.14}^{+0.17}$ & $1.27_{-0.13}^{+0.14}$ & $1.26_{-0.12}^{+0.15}$ \\ 
   260 & $1.38_{-0.17}^{+0.21}$ & $1.33_{-0.15}^{+0.18}$ & $1.36_{-0.14}^{+0.15}$ & $1.36_{-0.13}^{+0.16}$ \\ 
   300 & $1.48_{-0.18}^{+0.22}$ & $1.44_{-0.16}^{+0.19}$ & $1.46_{-0.15}^{+0.16}$ & $1.46_{-0.15}^{+0.16}$ \\ 
   350 & $1.59_{-0.19}^{+0.23}$ & $1.55_{-0.18}^{+0.20}$ & $1.57_{-0.16}^{+0.18}$ & $1.58_{-0.16}^{+0.18}$ \\ 
   400 & $1.69_{-0.20}^{+0.25}$ & $1.65_{-0.19}^{+0.22}$ & $1.67_{-0.17}^{+0.19}$ & $1.67_{-0.17}^{+0.19}$ \\  
\midrule \midrule 
\end{tabular}
\label{tab:esti_profile}
\end{table}

\section{Influence of a massive neighbor or satellite on halo mass estimate}\label{sec:massive_test}

As shown in \citet{Li2017, Li2019}, a massive neighbor or satellite has little effect
on the halo mass estimated from a simulation-based DF. Therefore, our estimated MW halo mass should be
insensitive to the presence of M31 and the LMC.
Here we demonstrate this insensitivity for the EAGLE DF with mock observations.
We only use those EAGLE halos that have at least 10 luminous satellites 
with $40<r<300\,\kpc$ per halo.
We estimate the mass of a test halo from the EAGLE DF with the prior based on the $M$-$c$ relation.
The results on the influence of the nearest more massive neighbor or the most massive satellite
are shown in \reffig{fig:test_massive}.

As discussed in the appendix of \citet{Li2017}, 
the relative strength of the external tidal field from a neighbor 
can be characterized by $(d_\mathrm{ngb}/R_\mathrm{ngb})^{-3}$, where
$d_\mathrm{ngb}$ is the distance to the neighbor and $R_\mathrm{ngb}$ is its virial radius.
We locate every more massive halo in the neighborhood of a test halo 
and define the one with the smallest $d_\mathrm{ngb}/R_\mathrm{ngb}$ as the nearest 
more massive neighbor.
As shown in the upper panel of \reffig{fig:test_massive}, the halo mass estimate is independent of 
this ratio. 
Note that $d_\mathrm{M31}/R_\mathrm{M31}\simeq 4$ for our MW.

The lower panel of \reffig{fig:test_massive} shows that the halo mass estimate is 
essentially independent of the subhalo mass $m_\mathrm{sat,max}$ of the most massive satellite when it is
below 1/20 of the host halo mass $M$.
There appears to be a very weak overestimate and a slightly larger scatter for $m_\mathrm{sat,max}/M>1/20$. 
Even for this case, the effects are much smaller than the statistical error.\footnote{
A much larger bias (up to 50\%) due to the LMC is reported in \citet{Erkal2020}. 
We note that a different mass estimator is used in their analysis
and the quoted bias would be much smaller 
if the estimated mass is compared with the total mass of the MW including the LMC.}
Nevertheless, it is worthwhile to quantify the effects more precisely with a larger halo sample in the future,
considering that the LMC might exceed 1/5 of the MW mass (e.g., \citealt{Penarrubia2016,Fritz2019}).

\begin{figure}[htbp]
  \centering
  \includegraphics[width=0.45\textwidth]{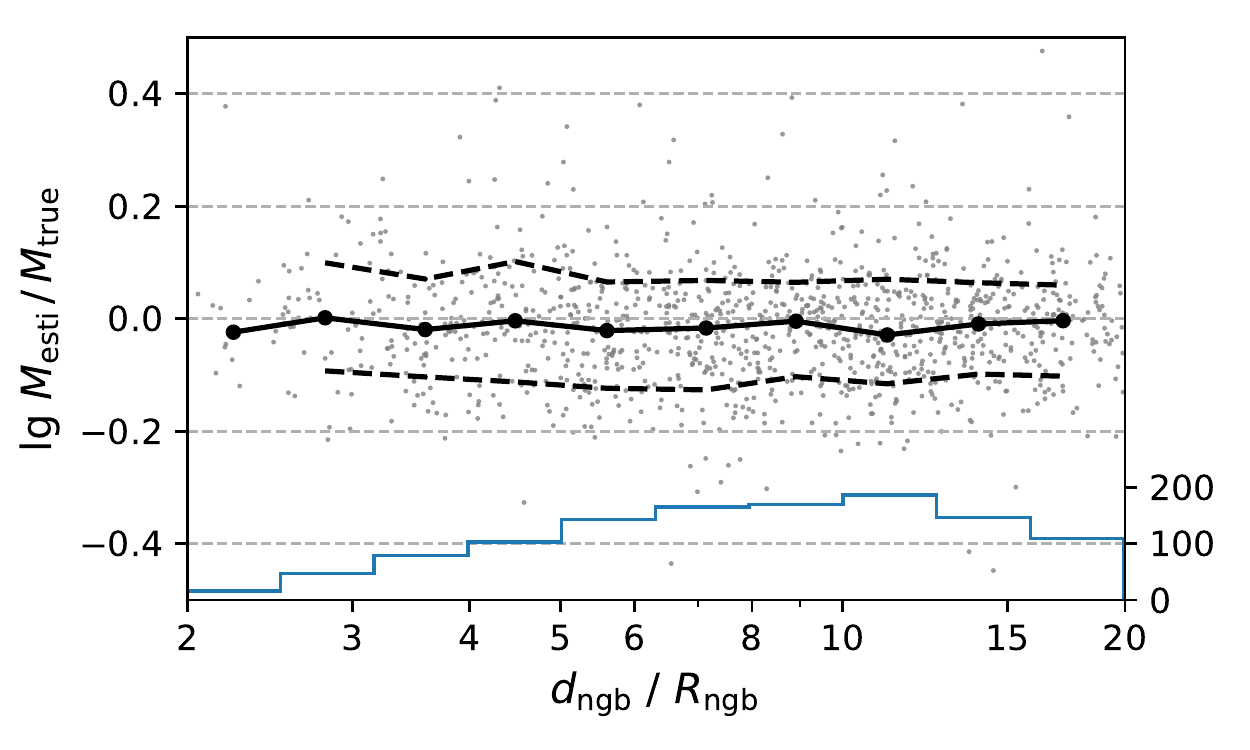}
  \includegraphics[width=0.45\textwidth]{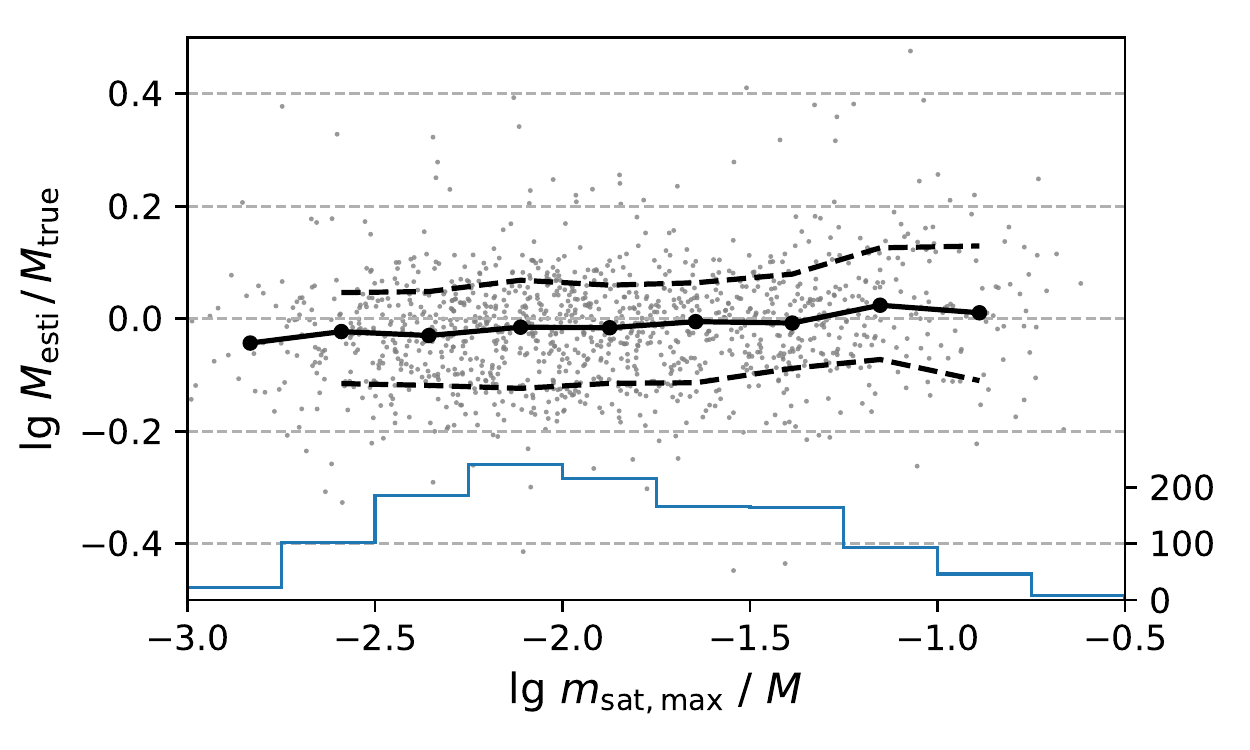}
  \caption{
  Influence of a massive neighbor (upper panel) or satellite (lower panel) on the halo mass estimate.
  Halo mass estimates from the EAGLE DF in terms of $\lg M_\mathrm{esti}/M_\mathrm{true}$ are shown as 
  gray dots. The histograms show the number of test halos in each bin.
  The solid and dashed black curves show the median and the $1\sigma$ interval of 
  $\lg M_\mathrm{esti}/M_\mathrm{true}$ for each bin. The nearest more massive neighbor is characterized 
  by the ratio between its distance $d_\mathrm{ngb}$ to the test halo and its virial radius $R_\mathrm{ngb}$. 
  The most massive satellite is characterized by the ratio between its subhalo mass $m_\mathrm{sat,max}$
  and the halo mass $M$. See text for details.
  }
  \label{fig:test_massive}
\end{figure}

\section{Enhanced satellite disruption and 
uncertainty in hydrodynamics-based simulations}\label{sec:hydro_esti}

\begin{figure*}[htbp]
  \centering
  \includegraphics[width=0.45\textwidth]{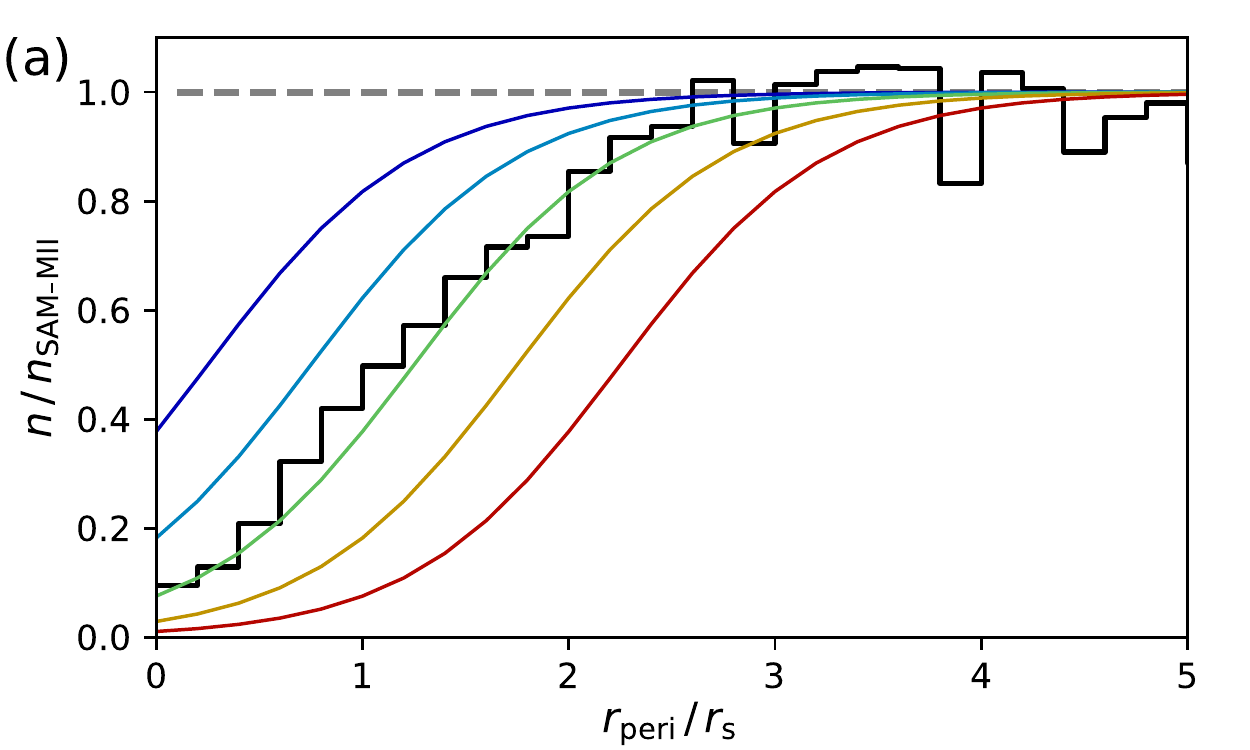}
  \includegraphics[width=0.45\textwidth]{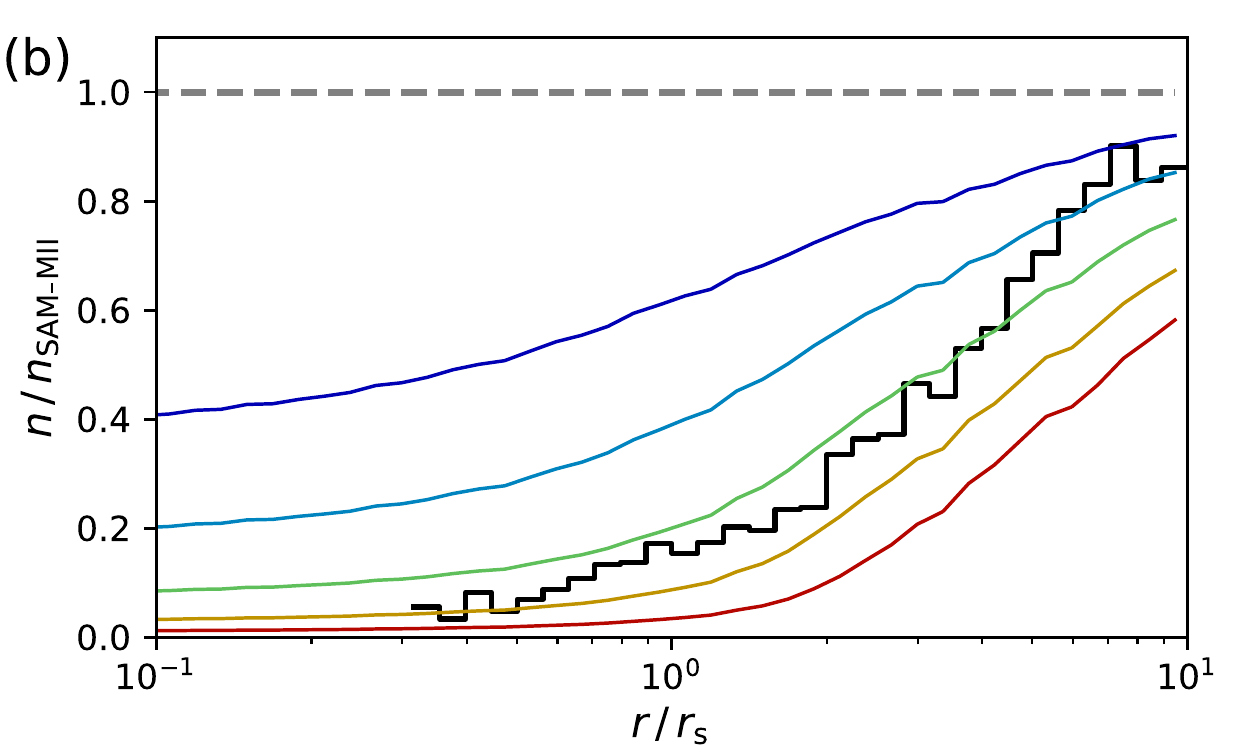}
  \includegraphics[width=0.45\textwidth]{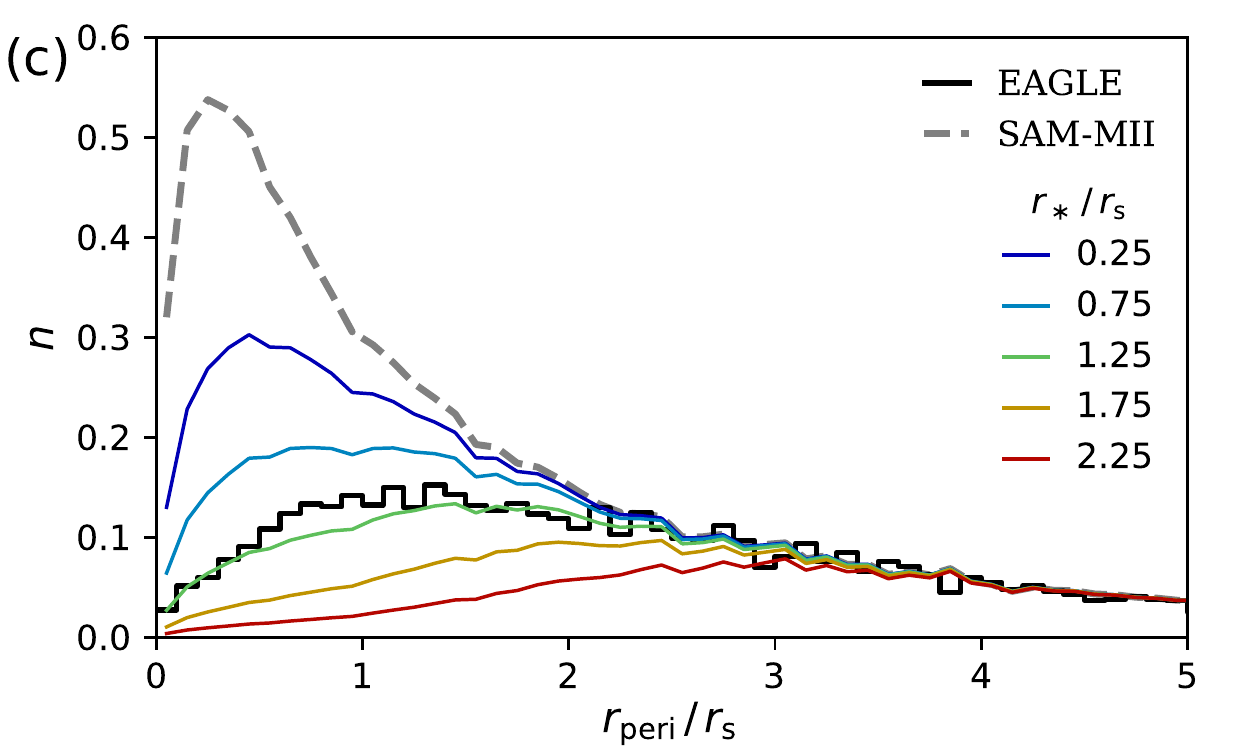}
  \includegraphics[width=0.45\textwidth]{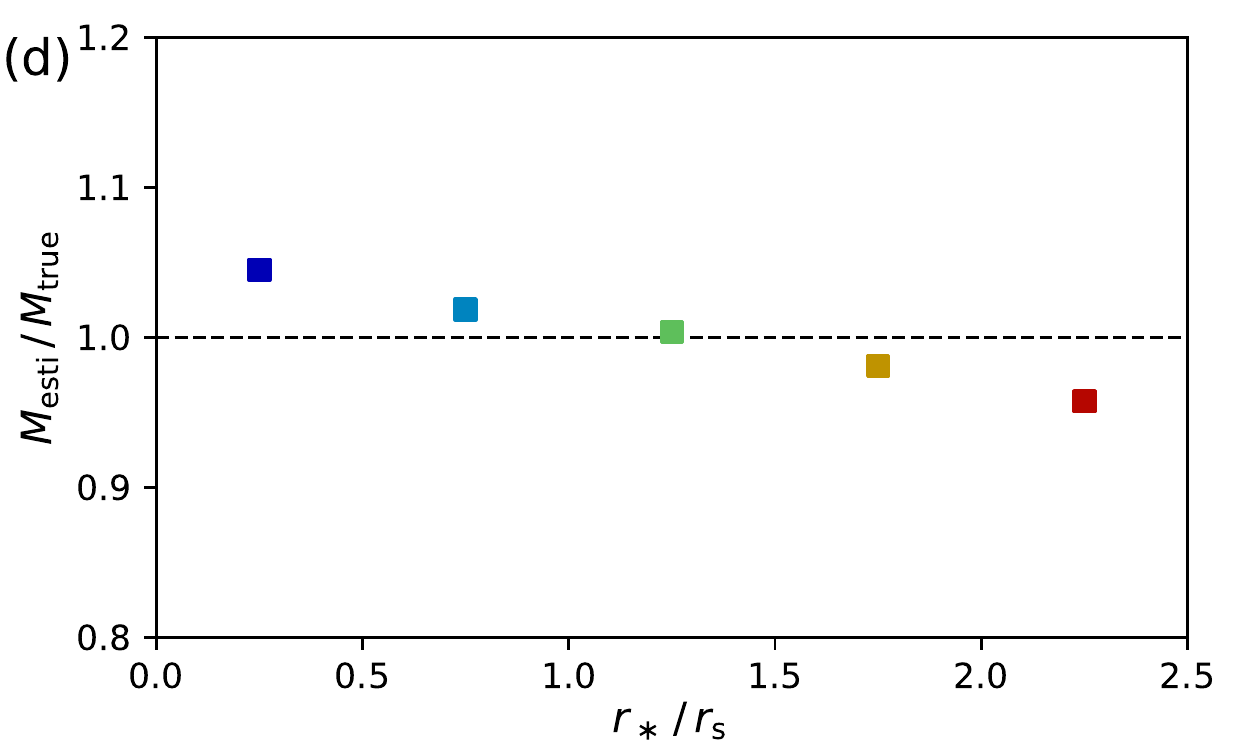}
  \caption{Influence of enhanced satellite disruption.
  The dashed gray and solid black curves represent the SAM-MII and EAGLE simulations, respectively,
  while the colored curves represent modified SAM-MII simulations with different enhancement of satellite 
  disruption.
  (a) Ratio of the numbers of satellites in the modified and original SAM-MII simulations
  as a function of $\rperi/\rs$ for different enhanced disruption prescribed by $r_\ast$.
  The value of $r_\ast=1.25\rs$ can approximately describe the EAGLE simulation.
  (b) Similar to (a) but for the radial distribution.
  (c) Similar to (a) but for the number of satellites as a function of $\rperi/\rs$.
  (d) Median mass estimates from modified SAM-MII DFs for mock observations of EAGLE halos.
  }
  \label{fig:destruction_illustrate}
\end{figure*}

While the hydrodynamics-based EAGLE simulation provides a better description of the observed
MW satellite kinematics than the SAM-MII simulation, variation in treatment of physical processes 
in current hydrodynamics-based simulations also leads to
scatter in the estimate of halo properties from the DF method.
Here we estimate this scatter by mimicking enhanced satellite disruption in the SAM-MII simulation.

The central galaxy potential in hydrodynamics-based simulations can enhance satellite disruption
in the inner halo, e.g., due to the enhancement of the tidal field by the stellar disk
\citep[e.g.,][]{Garrison-Kimmel2017,Kelley2018}.
We mimic this enhanced satellite disruption by manually removing
a fraction $1-F_\mathrm{surv}$ of satellites from each SAM-MII template halo. Here
\begin{equation}
    F_\mathrm{surv} = \frac{1}{1+\exp[2(r_\ast - \rperi)/\rs]},
\end{equation}
$r_\ast$ is the characteristic pericenter distance for which half of the satellites are disrupted,
and $\rs$ is the characteristic radius of the NFW halo profile.\footnote{
This prescription is only a simple approximation.
In addition to the pericenter distance, the disruption rate also depends on the apocenter distance.
Satellites with larger apocenter distances spend more time in the outer region and hence are less 
affected by the inner potential of the central galaxy.}
Satellites with $\rperi \gg r_\ast$ are not affected.
We can mimic different enhancement of disruption by varying $r_\ast$, with 
$r_\ast=-\infty$ corresponding to no enhanced disruption and larger $r_\ast$ 
to more enhanced disruption. Modified distributions of satellites for various $r_\ast$ are shown 
as functions of $\rperi$ and $r$ in \reffig{fig:destruction_illustrate} 
(see \citealt{Garrison-Kimmel2017} for similar figures).
The distributions for the EAGLE simulation can be approximated by $r_\ast=1.25\rs$.
Comparing \reffig{fig:destruction_illustrate} (b) and Figure 7 of \citet{Richings2018}, 
we estimate $r_\ast\sim 0.25\rs$ and $1.75\rs$ for the APOSTLE and Auriga simulations, respectively.
The central galaxies in the Auriga simulation are more massive and hence more efficient at
disrupting satellites.

Using the modified SAM-MII satellite samples, we construct the corresponding DFs and apply 
them to estimate halo properties with mock observations of EAGLE halos.
The prior based on the $M$-$c$ relation is used.
\reffig{fig:destruction_illustrate} (d) shows the dependence of the halo mass estimate on 
the enhancement of satellite disruption. As the $r_\ast$ for the modified SAM-MII DF changes from 
$1.25\rs$ approximating the EAGLE simulation to $0.25\rs$ ($1.75\rs$) approximating the APOSTLE 
(Auriga) simulation, the resulting systematic bias in the halo mass estimate is $\lesssim 5\%$,
which is negligible compared to the statistical uncertainty.
This result is not surprising. Although the number of satellites changes due to different
enhancement of disruption, their velocity distribution in the outer halo is less affected 
\citep[e.g.,][]{Sawala2017,Richings2018}. Because the halo mass estimate is mainly constrained by 
the velocity distribution rather than the spatial distribution \citep{Li2019}, this estimate
is insensitive to the differences among hydrodynamics-based simulations. The above result is
also consistent with the findings of \citet{Callingham2018}, who recovered halo masses in the
Auriga simulation with little bias using the orbital distribution from the EAGLE simulation.

\begin{figure}[htbp]
  \centering
  \includegraphics[width=0.45\textwidth]{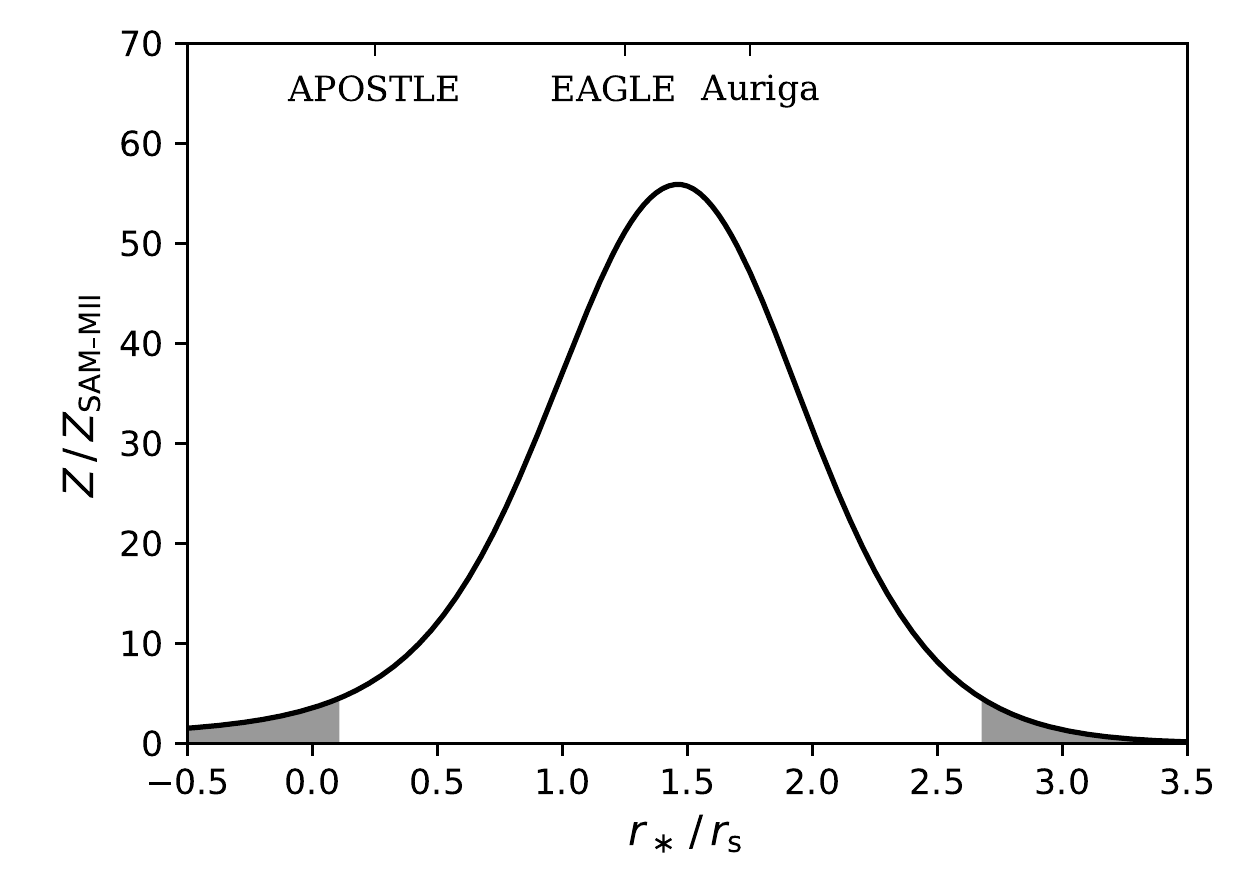}
  \caption{
  Constraints on enhanced satellite disruption from the observed MW satellite kinematics.
  The Bayesian factor of the modified relative to the original SAM-MII DF is shown as a function of $r_\ast/\rs$.
  Values of $r_\ast/\rs$ disfavored at the $2 \sigma$ level are indicated by the gray shades.
  Approximate values of $r_\ast/\rs$ for the hydrodynamics-based APOSTLE, EAGLE, and Auriga simulations
  are also indicated.
  }
  \label{fig:destruction_evidence}
\end{figure}

We calculate the Bayesian evidence of the modified SAM-MII DFs for the observed MW satellite kinematics.
The results are shown in \reffig{fig:destruction_evidence}. The DF with $r_\ast=1.5 \rs$ is the most 
favored. The corresponding projected DF $p_\mathrm{s}(r, \vt)$ is shown in \reffig{fig:destruction},
and indeed matches the observations very well. The values of $r_\ast$ approximating the hydrodyanmics-based
APOSTLE, EAGLE, and Auriga simulation suites are indicated in \reffig{fig:destruction_evidence}.
It can be seen that all three simulations are allowed by the current observations, 
though the APOSTLE results seem less favored (see also \citealt{Riley2018}).
Clearly, the above comparison of these simulations is indirect and approximate.
In the future, more actual simulations should be used to evaluate their Bayesian evidence as done for the
EAGLE simulation in this study. For more precise comparison of the simulations, it helps to have a larger 
sample of MW satellites with more accurate data or a stacked sample of galaxy groups or clusters.

\bibliography{MW_MassIV}

\end{document}